\documentclass[fleqn,usenatbib]{mnras}

\usepackage[T1]{fontenc}
\usepackage{ae,aecompl}
\usepackage[dvipsnames]{xcolor}
\usepackage{tabularx}

\usepackage{graphicx}	% Including figure files
\usepackage{amsmath}	% Advanced maths commands
\usepackage{amssymb}	% Extra maths symbols
\usepackage{natbib}
\usepackage{color}
\usepackage{subfigure}

\def\caii{\relax \ifmmode {\mbox Ca\,{\scshape ii}}\else Ca\,{\scshape ii}\fi}

\title[Leo~I stellar content]{Dissecting the stellar content of Leo~I: a dwarf irregular caught in transition}

\author[T. Ruiz-Lara et al.]{T. Ruiz-Lara$^{1, 2, 3}$\thanks{E-mail: tomasruizlara@gmail.com (TRL)}, C. Gallart$^{2, 3}$,  M. Monelli$^{2, 3}$, T. K. Fritz$^{2, 3}$, G. Battaglia$^{2, 3}$, \newauthor S. Cassisi$^{4, 5}$, M. Luis Aznar$^{2}$, A. V. Russo Cabrera$^{2,6}$, I. Rodr\'iguez-Mart\'in$^{7}$, \newauthor J. J. Salazar-Gonz\'alez$^{7}$ \\
$^{1}$ Kapteyn Astronomical Institute, University of Groningen, Landleven 12, 9747 AD Groningen, The Netherlands \\
$^{2}$ Instituto de Astrof\'isica de Canarias, Calle V\'ia L\'actea s/n, E-38205 La Laguna, Tenerife, Spain \\
$^{3}$ Departamento de Astrof\'isica, Universidad de La Laguna, E-38200 La Laguna, Tenerife, Spain \\
$^{4}$ INAF -- Astronomical Observatory of Abruzzo, via M. Maggini, sn, 64100 Teramo, Italy \\
$^{5}$ INFN, Sezione di Pisa, Largo Pontecorvo 3, 56127 Pisa, Italy  \\
$^{6}$ Grupo de Operaciones de Ingenier\'ia, Gran Telescopio de Canarias, Cuesta de San Jos\'e, s/n, E-38712 Bre\~na Baja, La Palma, Spain \\
$^{7}$ Dpto. de Matem\'aticas, Estad\'istica e Investigaci\'on Operativa, Universidad de La Laguna, E-38200 La Laguna, Tenerife, Spain
}

\date{Accepted 2020 November 24. Received 2020 November 24; in original form 2020 October 13}

\pubyear{2020}

\begin{document}
\label{firstpage}
\pagerange{\pageref{firstpage}--\pageref{lastpage}}
\maketitle

\begin{abstract}

Leo~I is considered one of the youngest dwarf spheroidals (dSph) in the Local Group. Its isolation, extended star formation history (SFH), and recent perigalacticon passage ($\sim$1~Gyr ago) make Leo~I one of the most interesting nearby stellar systems. Here, we analyse deep photometric Hubble Space Telescope data via colour-magnitude diagram fitting techniques to study its global and radially-resolved SFH. We find global star formation enhancements in Leo~I $\sim$13, 5.5, 2.0, and 1.0~Gyr ago, after which it was substantially quenched. Within the context of previous works focused on Leo~I, we interpret the most ancient and the youngest ones as being linked to an early formation (surviving reionisation) and the latest perigalacticon passage (transition from dIrr to dSph), respectively. We clearly identify the presence of very metal poor stars ([Fe/H]$\sim$-2) ageing $\sim$5-6 and $\sim$13~Gyr old. We speculate with the possibility that this metal-poor population in Leo~I is related to the merging with a low mass system (possibly an ultra-faint dwarf). This event would have triggered star formation (peak of star formation $\sim$5.5~Gyr ago) and accumulated old, metal poor stars from the accreted system in Leo~I. Some of the stars born during this event would also form from accreted gas of low-metallicity (giving rise to the 5-6~Gyr low-metallicity tail). Given the intensity and extension of the 2.0~Gyr burst, we hypothesise that this enhancement could also have an external origin. Despite the quenching of star formation around 1~Gyr ago (most probably induced by ram pressure stripping with the Milky Way halo at pericentre), we report the existence of stars as young as 300-500 Myr. We also distinguish two clear spatial regions: the inner $\sim$190~pc presents an homogeneous stellar content (size of the gaseous star forming disc in Leo~I from $\sim$4.5 to 1~Gyr ago), whereas the outer regions display a clear positive age gradient.

\end{abstract}


%%%%%%%%%%%%%%%%%%%%%%%%%%%%%%%%%%%%%%%%%%%%%%%%%%

%%%%%%%%%%%%%%%%% BODY OF PAPER %%%%%%%%%%%%%%%%%%

\section{Introduction}

\newpage

The interest on galaxies in the low mass end of the galaxy population has been vigorous in the last decades \citep[e.g. ][]{1998ARA&A..36..435M, 1999AJ....118.2245G, 2006A&A...459..423B, 2007ApJ...654..897B, 2009ARA&A..47..371T, 2010ApJ...720.1225M, 2011ApJ...739....5W, 2011ApJ...730...14H, 2011MNRAS.415L..40B, 2012AJ....144....4M, 2013ApJ...779..102K}. This raising interest is well founded. On one hand, dwarf galaxies, the most common kind in the Universe, have masses in rough agreement with the critical mass of the first galaxies that turned into the building blocks of the stellar haloes of present-day massive galaxies (e.g.  \citealt[][]{1965ApJ...142.1317P}; see \citealt[][]{2018PhR...780....1D} for a review). Indeed, some of the dwarf galaxies that we currently observe in the Local Group could be analogues of the first dwarf galaxies to exist in the Universe and are survivors of such an early epoch. On the other hand, the study of dwarf galaxies is crucial to improve cosmological models of galaxy formation on small scales, where baryon physics involved in star formation, inflows and outflows or stellar feedback need to be properly described \citep[e.g.][]{2012RAA....12..917S, 2014MNRAS.437..415D, 2016MNRAS.461.2658D}. As such, the characterisation of the stars populating these systems as well as of their star formation histories\footnote{In this work we will consider the star formation history of a stellar system as the evolution with the cosmic time of both the star formation rate (SFR) and the chemical enrichment (Age-Metallicity Relation, AMR).} (SFH), is of the utmost importance to understand, not only dwarf galaxy evolution \citep{2009ARA&A..47..371T, 2015ApJ...811L..18G}, but the evolution of the Universe as a whole from its earliest epochs \citep[e.g.][]{2014ApJ...789..148W, 2015MNRAS.450.4207B, 2016ApJ...819..147M}.

Dwarf galaxies have been classified historically in dwarf spheroidal galaxies (dSph), transition types (dT) or irregular (dIrr), mainly based on their {\it current} morphology as observed in optical bands (from smooth and spheroidal to patchy and irregular); and gas and young stellar content (absence to richness). It seems natural that, based on this classification, dwarf galaxies evolve with time from dIrr to dT and finally to dSph. In this process, the environment where they evolve should play an important role as well \citep[as indicated by the morphology-density relation found in the Local Group,][]{1999IAUS..192...17G}. However, a classification based on their complete SFH (actual evolution) should provide a more complete and precise view on how these systems form and evolve. In this line, \citet[][]{2015ApJ...811L..18G}, using data from the LCID project\footnote{\url{http://research.iac.es/proyecto/LCID/}}, proposed a new classification of dwarf galaxies as ``{\it slow}'' or ``{\it fast}''. {\it Fast} dwarfs would be composed mainly by old stars, outcome of a short and intense early star formation, whereas {\it slow} dwarfs would only form a small fraction of their stellar mass at early epochs, presenting an extended SFH running up to (nearly) the present time. Taking into account the location and inferred orbits of the studied dwarfs, \citet[][]{2015ApJ...811L..18G} proposed that the ultimate origin of the dychotomic SFHs, and thus of the two dwarf galaxy types, would be related to the density of the environment where each dwarf formed, being the ``{\it slow}'' or ``{\it fast}'' {\it nature} imprinted at formation. Later evolution (inflows, outflows, dwarf-dwarf interactions, interaction with host galaxy, feedback, etc.) would further delineate the evolution and current properties of ``{\it slow}'' dwarfs. This includes a possible environmentally induced late star formation quenching that could give a ``{\it slow}'' dwarf, with an evolutionary history much alike that of a classical dIrr galaxy, the current appearance of a classical dSph. As a consequence, the analysis of the stellar content of {\it slow} dwarfs, identified as those with substantial star formation beyond 10 Gyr or so, is of special importance.

Having formed stars since the very beginning up to advanced stages in the evolution of the Universe, the study of dwarf galaxies with extended SFHs can help us in understanding the complex interplay between halo growth, reionization, star formation, stellar feedback, and environmental effects. When and how these galaxy evolution drivers affect dwarf galaxies depend on the specifics of each system at different stages of its evolution, and imprint different signatures in their SFHs. In particular, it is known that this evolution can induce internal spatial variations of the stellar content characteristics in dwarf galaxies. 

Dwarf galaxies tend to be younger and more metal-rich in their inner parts than in their outskirts, where stars are older and more metal-poor \citep[e.g.][]{2006A&A...459..423B, 2007A&A...465..357F, 2014A&A...570A..78B, 2015MNRAS.454.3996D, 2017MNRAS.467..208O}. These stellar populations gradients have been shown to exist since very early epochs, from the radial variation of the properties of the old RR Lyrae variable stars in ``{\it fast}'' galaxies \citep[e.g.][]{Bernard2008Tucana, Martinez-Vazquez2015Sculptor}.   Their existence can be explained as a consequence of the shrinking of the star forming disc in dwarf galaxies, radial migration, and ejection of stars via supernova feedback \citep[][]{2009MNRAS.395.1455S, 2019MNRAS.490.1186G}. Indeed, \citet[][]{2019MNRAS.490.1186G} found, making use of cosmological simulations, that dwarf galaxies with large dark matter cores display flatter age profiles as a consequence of the effect of late-time feedback (extended SFH), predicting clear relations between radial changes in the SFH of dwarf galaxies and their dark matter halo distribution. Unfortunately, precise analysis of the spatial variation of extended SFH in dwarf galaxies exist only for a handful of systems \citep[e.g.][]{2013ApJ...778..103H}.

Leo~I is one of the most distant Milky Way satellites, and it is classified as a dSph galaxy \citep[][]{2012AJ....144....4M} based on its current morphology and lack of gas \citep[][]{1978AJ.....83..360K, 2009ApJ...696..385G}, or as a ``{\it slow}'' galaxy based on its full SFH \citep{2015ApJ...811L..18G}. Indeed, Leo~I displays a very extended SFH \citep[][]{1999AJ....118.2245G, 2000MNRAS.317..831H}, and is considered one of the youngest dSph in our Local Group in terms of average age \citep[see][]{1993AJ....106.1420L} together with Leo~A \citep[][]{2007ApJ...659L..17C, 2018A&A...617A..18R} or the Aquarius dwarf irregular galaxy \citep[DDO210][]{2014ApJ...795...54C}. Indeed, its oldest population remained elusive for some time \citep[][]{2000ApJ...530L..85H} and most of Leo~I star forming activity happened between 7 and as recently as 1 Gyr ago \citep[][]{1999AJ....118.2245G}. At such point, Leo~I stopped forming stars probably linked to the pericentric approach to the Milky Way $\sim$1 Gyr ago, as indicated by the careful analysis of the proper motion of Leo~I stars and derived orbit \citep[][]{2013ApJ...768..139S}. The dating of this pericentric passage of Leo~I has been recently confirmed using Gaia DR2 \citep[][]{2018A&A...616A...1G} data \citep[][]{2018A&A...616A..12G, 2018A&A...619A.103F}. Thus, it seems that Leo~I has just been caught in transition between dIrr to dSph. But this is just its most recent history. Previous works have hinted a turbulent past characterised by alternated periods of intense star formation with quiescent intervals \citep[][]{1999AJ....118.2245G, 2000MNRAS.317..831H}. All this, combined with the narrow stellar metallicity displayed by Leo~I, its shallow negative metallicity gradient and positive age trend \citep[][]{2009A&A...500..735G}, make of Leo~I one of the most exciting dwarf galaxies in the Local Group. However, most of the efforts at characterising the SFH of Leo~I made used of old observations from the WFPC2 on board of the Hubble Space Telescope (HST), dating more than 2 decades back \citep[][]{1999AJ....118.2245G, 2000MNRAS.317..831H, Dolphin2002, 2014ApJ...789..148W, 2015ApJ...804..136W}. A whole new characterisation of its SFH, with updated and deeper data and powerful analysis techniques will allow us to fully unveil the sequence of events shaping Leo~I, improving the age resolution of previous works and determining precise chemical enrichment histories. A summary of the main properties of this system can be found in Table~\ref{tab:leoi_props}.

\begin{table}
{\normalsize
\centering
\begin{tabular}{lcc}
\hline\hline
Parameter & Value & Reference \\ 
$(1)$ & $(2)$ & $(3)$ \\ \hline
R.A. (J2000)  &  10$^{\rm h}$08$^{\rm m}$28$^{\rm s}$.1   &  i \\ 
Decl. (J2000) &  +12$^{\rm o}$18$'$23$''$    & i  \\ 
l  &  226.0   &  i \\ 
b &  +49.1    & i  \\ 
Distance (kpc)  &   254   &  ii \\ 
(m-M)$_{\rm 0}$   &   22.02   &  ii \\ 
M$_{\rm V}$  &   -12.0 $\pm$ 0.3   &  iii \\ 
M$_{\rm *}$ (M$_\odot$)  & 5.5$\times$10$^{\rm 6}$  &  i \\ 
M$_{\rm *}$/L$_{\rm V}$    &   4.4   &  iv \\ 
M$_{\rm HI}$ (M$_\odot$)    &   <0.03$\times$10$^{\rm 6}$   &  v \\ 
M$_{\rm dyn}$ (R$_{\rm vir}$) (M$_\odot$)   &   (7 $\pm$ 1)$\times$10$^8$  &  vi \\ 
V$_{\rm sys}$ (km s$^{-1}$)  &   282.9 $\pm$ 0.5 &  vi \\ 
E(B-V)  &   0.036   &  i \\ 
R$_{\rm vir}$ (kpc) & 18.3 $\pm$ 2.7 & vi \\
R$_{\rm tidal}$ (pc) & 935 $\pm$ 140 & iii \\
R$_{\rm 1/2 light}$ (pc)  &   251 $\pm$ 27   &  iii \\ 
Ellipticity   &  0.21   &  iii \\ 
Position angle ($^{\rm o}$)  &   79   &  iii \\ 
$\langle$[Fe/H]$\rangle$ & -1.43 $\pm$ 0.01 & vii \\
\hline
\end{tabular}
\caption{Leo~I main properties. References: i-~\citet[][]{2012AJ....144....4M}; ii-~\citet[][]{2004MNRAS.354..708B}; iii-~\citet[][]{1995MNRAS.277.1354I}; iv-~\citet[][]{2018ApJ...860...66M}; v-~\citet[][]{2009ApJ...696..385G}; vi-~\citet{2008ApJ...675..201M}; vii-~\citet{2011ApJ...727...78K}.} 
\label{tab:leoi_props}
}
\end{table}

With this goal in mind, in this work we use novel colour-magnitude diagram (CMD) fitting techniques to analyse  a modern and extremely deep HST Advanced Camera for Surveys (ACS) dataset, reaching well below the oldest main sequence turnoff with  an unprecedented precise photometry. The combination of updated techniques and stellar evolutionary models, together with this exquisite dataset, will indeed improve the age resolution from previous works and to characterise the chemical enrichment of Leo~I as never before. Here, we revise the global SFH of Leo~I, adding important new insights as well as determine the radially-resolved SFH of Leo~I. This paper is structured as follows: Section~\ref{sec:data} describes the data as well as the data reduction. A qualitative analysis of the observed CMD of Leo~I is provided in Sect.~\ref{leoi_cmd_desc} whereas a quantitative study of its SFH (globally and radially) is given in Sects.~\ref{all_leoi} and~\ref{radial_leoi}. The results are discussed in Sect.~\ref{discussion} and the conclusions outlined in Sect.~\ref{conclusions}.

\section{Data sets and reduction process}
\label{sec:data}

\begin{figure*}
\centering 
\includegraphics[width = 0.95\textwidth]{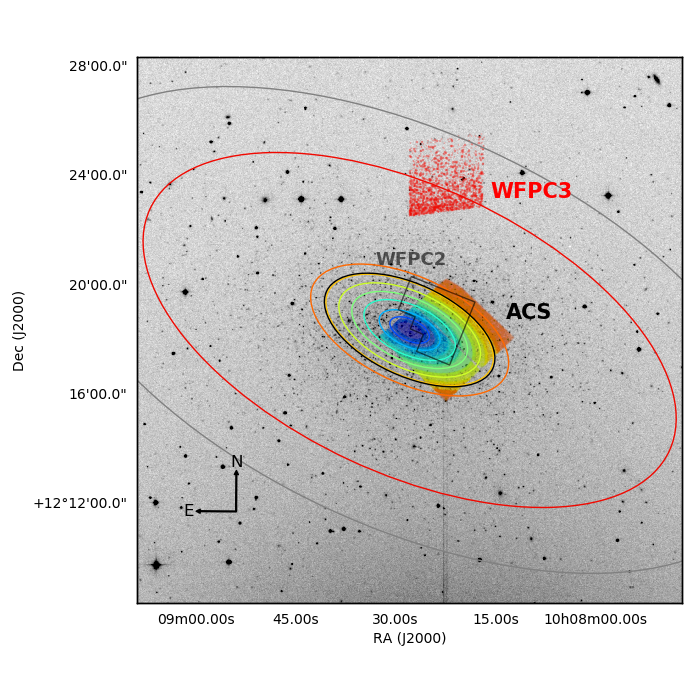} \\   
\caption{Leo~I observational layout. Spatial distribution of stars analysed in this work, colour-coded according to its radial distance (see Table~\ref{elliptical_tab}), superimposed to a DSS r-band image of the Leo~I dwarf galaxy. The main ACS field, together with the paralel WFC3 field, can be easily outlined (see Sect.~\ref{sec:data}). The black and grey ellipses depict the half-light and tidal radius, respectively. For completeness, we have added the footprint (grey polygon) of the WFPC2@HST data analysed in previous determinations of the SFH of Leo~I \citep[i.e.][]{1999AJ....118.2245G, 2000MNRAS.317..831H, Dolphin2002, 2014ApJ...789..147W, 2014ApJ...789..148W}.} 
\label{leoi_layout} 
\end{figure*}

Images of Leo~I were retrieved from the HST archive, and were originally collected by two projects: GO10520\footnote{\textit{Resolving the Complex Star Formation History of the Leo I Dwarf Spheroidal Galaxy}, P.I. T. Smecker-Hane}, and GO12270\footnote{\textit{Proper Motion of Leo I: Constraining the Milky Way Mass}, P.I. S. Sohn}. In both cases, the ACS was pointing the innermost region, though slightly off-centred, while the parallel WFC3 were located to the north (see Fig.~\ref{leoi_layout}). In the case of the ACS (WFC3), a total integration time of 10,200 (8,091) s and 27,297 (8,491) s, was devoted to the $F435W$ ($F438W$) and $F814W$ filters, respectively. Detailed logs are presented in Table~\ref{tab:log}.

\subsection{Data reduction}
\label{sec:reduction}

The photometric reduction was performed using the DAOPHOTIV and ALLFRAME suite of programs \citep{stetson87, stetson94}, using the same approach for both the ACS and WFC3 data \citep[see][for a detailed description of the procedure]{2010ApJ...720.1225M}. Individual empirical point spread functions (PSFs) were derived for individual chip of each image, with semi-automatic iterative routines. The input list of stars for ALLFRAME was generated on a median, cosmic-rays cleaned stacked image. ALLFRAME fits the empirical PSFs to candidate sources on individual  images, producing a catalogue of positions and  instrumental magnitudes for each image. The final list of sources was obtained cross-matching individual files, retaining objects detected in at least ($half+1$) of the images. The final photometric lists contain 191,756 and 8,956 sources for the ACS and the WFC3 fields. Instrumental magnitudes were calibrated to the VEGAMAG system using appropriate zero-points available on each instrument web page.

Artificial stars tests (ASTs) were performed following \citep[following][]{2010ApJ...720.1225M}. The 2$\times$10$^6$ stars injected in each chip, in many runs with few thousand stars each were drawn from a synthetic CMD covering well the observed one, and similar to the one that will be used to derive the SFH. Artificial stars were distributed in a regular grid and spaced avoiding overlap of the PSF's wings. The photometry was repeated identically as for the original data.  ASTs provide fundamental information to simulate the photometric errors in the synthetic CMD used to derive the SFH (see Sect.~\ref{all_leoi}).

As already commented, previous determinations of the SFH of Leo~I \citep[][]{1999AJ....118.2245G, 2000MNRAS.317..831H, Dolphin2002, 2014ApJ...789..148W, 2015ApJ...804..136W}, all made use of old data collected using the WFPC2@HST. For a complete account of such observations we encourage the reader to check \citet[][]{1999ApJ...514..665G}. A quick comparison between both datasets is outlined here. WFPC2 observations used filters F555W and F814W, whereas the ACS dataset analysed here uses filters F435W and F814W. The broader colour baseline of the data analysed in this work plays an important role at separating different stellar populations using CMD fitting. In addition, ACS observations clearly have larger exposure times as compared to the WFPC2 data (10,200 and 27,297 seconds compared to 6,300 and 5,100 seconds, blue and red filters, respectively). This ends up with our CMDs being populated by 191,756 stars compared to the 28,000 stars with reliable photometry from WFPC2 (as can be seen from Fig.~\ref{leoi_layout}, the old WFPC2 field is almost fully embedded within the new ACS field). Finally, while WFPC2 data hardly reaches oMSTO \citep[near I-band 26 mag according to][]{1999ApJ...514..665G}, the CMD presented in Fig.~\ref{leoi_cmd} reaches well below the oMSTO, and down to F814W $\sim$ 5 mag with reliable photometry (40\% completeness according to our ASTs). We have shown in \citet[][]{2018A&A...617A..18R} that the analysis of CMDs of a similar depth to the ones observed with WFPC2 present clear disadvantages with respect to deeper CMDs when recovering SFHs, especially at old ages.

The improvements of the here-analysed photometric dataset and our analysis will allow us to achieve an unprecedented age resolution and AMR reconstruction. The coupling of all these factors will result on a unique opportunity to pinpoint key events and fully unveil the past history of Leo~I.

\begin{table}
\scriptsize
\centering
\begin{tabular}{lcccc}
\hline\hline
  Name & Filter & Exp. time & UT start & Prog. ID  \\ 
       &        &    (s) &  &   \\ 
  (1)     &  (2)      &    (3) & (4) & (5)  \\ 
\hline
\multicolumn{5}{c}{ACS data} \\
\hline
j9gz01orq & $F435W$ & 1700.0 & 2006-02-10 09:38:31 & 10520 \\
j9gz01oxq & $F435W$ & 1700.0 & 2006-02-10 14:22:51 & 10520 \\
j9gz02tiq & $F435W$ & 1700.0 & 2006-02-18 06:18:54 & 10520 \\
j9gz02tjq & $F435W$ & 1700.0 & 2006-02-18 07:50:23 & 10520 \\
j9gz02tlq & $F435W$ & 1700.0 & 2006-02-18 09:26:19 & 10520 \\
j9gz02tnq & $F435W$ & 1700.0 & 2006-02-18 11:02:15 & 10520 \\
j9gz03tqq & $F435W$ & 1700.0 & 2006-02-18 12:39:33 & 10520 \\
j9gz04tsq & $F814W$ & 1700.0 & 2006-02-18 14:17:56 & 10520 \\
j9gz04ttq & $F814W$ & 1700.0 & 2006-02-18 15:50:04 & 10520 \\
j9gz04tvq & $F814W$ & 1700.0 & 2006-02-18 17:26:01 & 10520 \\
j9gz05tyq & $F814W$ & 1700.0 & 2006-02-18 19:06:10 & 10520 \\
j9gz05tzq & $F814W$ & 1700.0 & 2006-02-18 20:41:09 & 10520 \\
j9gz05u1q & $F814W$ & 1700.0 & 2006-02-18 22:18:22 & 10520 \\
j9gz06krq & $F814W$ &  440.0 & 2006-01-26 06:35:01 & 10520 \\
j9gz06ksq & $F814W$ &  440.0 & 2006-01-26 06:45:01 & 10520 \\
j9gz06kuq & $F814W$ &  440.0 & 2006-01-26 08:07:31 & 10520 \\
jbjm01kkq & $F814W$ & 1267.0 & 2011-01-25 11:46:18 & 12270 \\
jbjm01kmq & $F814W$ & 1267.0 & 2011-01-25 12:10:03 & 12270 \\
jbjm01kpq & $F814W$ & 1361.0 & 2011-01-25 13:18:10 & 12270 \\
jbjm01ktq & $F814W$ & 1360.0 & 2011-01-25 13:43:29 & 12270 \\
jbjm02kyq & $F814W$ & 1267.0 & 2011-01-25 14:58:04 & 12270 \\
jbjm02l0q & $F814W$ & 1267.0 & 2011-01-25 15:21:49 & 12270 \\
jbjm02l3q & $F814W$ & 1363.0 & 2011-01-25 16:29:57 & 12270 \\
jbjm02l7q & $F814W$ & 1364.0 & 2011-01-25 16:55:18 & 12270 \\
jbjm03lcq & $F814W$ & 1267.0 & 2011-01-25 18:09:50 & 12270 \\
jbjm03leq & $F814W$ & 1267.0 & 2011-01-25 18:33:35 & 12270 \\
jbjm03lhq & $F814W$ & 1363.0 & 2011-01-25 19:41:43 & 12270 \\
jbjm03llq & $F814W$ & 1364.0 & 2011-01-25 20:07:04 & 12270 \\
\hline
\multicolumn{5}{c}{WFC3 data} \\
\hline
ibjm01klq & $F814W$ & 1326.0 & 2011-01-25 11:45:23 & 12270 \\
ibjm01knq & $F438W$ & 1367.0 & 2011-01-25 12:10:24 & 12270 \\
ibjm01kqq & $F814W$ & 1371.0 & 2011-01-25 13:18:25 & 12270 \\
ibjm01kuq & $F438W$ & 1466.0 & 2011-01-25 13:43:50 & 12270 \\
ibjm02kzq & $F814W$ & 1326.0 & 2011-01-25 14:57:09 & 12270 \\
ibjm02l1q & $F438W$ & 1367.0 & 2011-01-25 15:22:10 & 12270 \\
ibjm02l4q & $F814W$ & 1371.0 & 2011-01-25 16:30:12 & 12270 \\
ibjm02l8q & $F438W$ & 1462.0 & 2011-01-25 16:55:39 & 12270 \\
ibjm03ldq & $F814W$ & 1326.0 & 2011-01-25 18:08:55 & 12270 \\
ibjm03lfq & $F438W$ & 1367.0 & 2011-01-25 18:33:56 & 12270 \\
ibjm03liq & $F814W$ & 1371.0 & 2011-01-25 19:41:58 & 12270 \\
ibjm03lmq & $F438W$ & 1462.0 & 2011-01-25 20:07:25 & 12270 \\
\hline
\end{tabular}
\caption{Log of the observations. (1) Image name; (2) name of the filter; (3) exposure time; (4) time (UT) of the start of the exposure; (5) identifier of the observing program.}
\label{tab:log}
\end{table}

\section{The Leo~I Colour-Magnitude Diagram}
\label{leoi_cmd_desc}

\begin{figure*}
\centering 
\includegraphics[width = 0.95\textwidth]{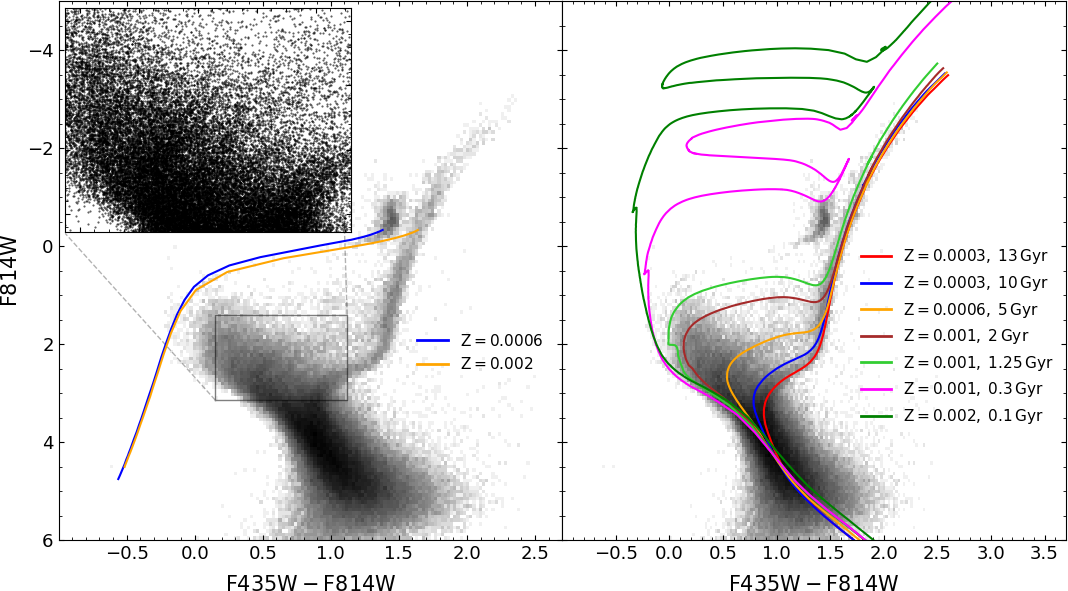} \\   
\caption{Observed M$_{\rm F814W}$ and M$_{\rm F435W}$-M$_{\rm F814W}$ colour-magnitude diagram from the ACS field in Leo~I. Left panel: Observed CMD with the Zero-Age Horizontal Branch (ZAHB) loci for metallicities of 0.002 and 0.0006 ([Fe/H]$\sim$-0.96 and -1.49, respectively), overlaid. To highlight the level of details that can be observed in the diagram, in particular the different and well-differentiated sub-giant branches departing from the main sequence, this panel includes an inset zooming in the region of the MSTOs. Right panel: This time we overlay a set of theoretical isochrones to the observed diagram. Ages and metallicities of these isochrones range from 100 Myr to 13 Gyr and from Z=0.002 ([Fe/H]$\sim$-0.96) to Z=0.0003 ([Fe/H]$\sim$-1.78), respectively. Adopted distance modulus and reddening as provided in Table~\ref{tab:leoi_props}.} 
\label{leoi_cmd} 
\end{figure*}

Figure~\ref{leoi_cmd} shows the derived ACS ($F814W$, $F435W-F814W$) CMD, which
spans over 9 magnitudes, from the tip of the red giant branch (RGB, $F814W$ $\sim$~-3 mag) down to about 2 mag fainter than the oldest main sequence (MS) turn-off (oMSTO), and it reveals many features strongly suggesting a complex SFH. The well populated faintest sub-giant branch (SGB) and RGB suggest an important fraction of old population, supported by the clear detection of a blue horizontal branch (HB). Nevertheless, the MS extends as a conspicuous sequence to bluer colour and brighter magnitude, indicating, together with the well populated red-clump, the presence of an important intermediate-age population. Interestingly, brighter than the old SGB, at least two other  SGB are clearly visible, separated by regions with lower density of points, suggesting a gasping SFH, with alternating periods of enhanced and more quiet star formation. Finally, the blue MS stars, reaching above the HB level, together with the presence of blue-loop stars, suggest the presence of a small population younger than 1~Gyr.

To further constrain the properties of the stellar populations in Leo~I, Fig.~\ref{leoi_cmd} presents the superposition of model predictions  
from the BaSTI database\footnote{\url{http://basti.oa-teramo.inaf.it/index.html}} \citep[][the same used for the quantitative derivation of SFH, see Sect.~\ref{all_leoi}]{2004ApJ...612..168P}. The left panel shows the superposition of the Zero-Age Horizontal Branch (ZAHB) for two metallicities, Z=0.0006, 0.002. The two sequences bracket the luminosity the HB stars over the whole colour range, suggesting a metallicity spread. The right panel presents the superposition of isochrones
for the labelled ages and metallicities. The comparison suggests the presence of an old ($\sim$10-13 Gyr) and relatively metal-poor (Z=0.0003, [Fe/H]$\sim$-1.8) population. The second SGB is well represented by a 5 Gyr, sligthly more metal-rich isochrone, while the upper SGB is compatible with a 2 Gyr, Z=0.001 ([Fe/H]$\sim$-1.2) population. Few tenths of magnitude below the luminosity level of the HB, the MS seems to present a quite sharp drop in stellar counts, suggesting a decrease in the star formation rate around 1 Gyr ago (green isochrone). Nevetheless, the magenta isochrone represents well the upper MS and the blue loop with a population of $\sim$300 Myr, although the dark green one shows that we cannot exclude the presence of some stars as young as 100 Myr.

Such a comparison, while enlightening about the Leo~I stellar content, suffers from two main problems: {\em i)} it is qualitative, in the sense that it can identify the presence of stellar populations, but it does not provide a quantitative estimate of the amount of mass formed; {\em ii)} it suffers from degeneracy between age and metallicity, such that isochrones with different combinations of these parameters can represent certain features in the CMD equally well. To overcome these problems, a detailed quantitative SFH is derived in Sect.~\ref{all_leoi}.

\section{The global Leo~I Star Formation History}
\label{all_leoi}

The quantitative analysis of the SFH of extended regions within dwarf galaxies is crucial to understand the past history of these systems, dating and characterising their main evolutionary events. In this section we derive the SFH of the central region of Leo~I (ACS field, see Fig.~\ref{leoi_layout}), while the possible spatial differences will be discussed in Section \ref{radial_leoi}. The larger number of stars used for the global analysis enables the proper assessment of the reliability and robustness of the SFH recovery method that is presented in Appendix A.

\begin{figure*}
\centering 
\includegraphics[width = 0.97\textwidth]{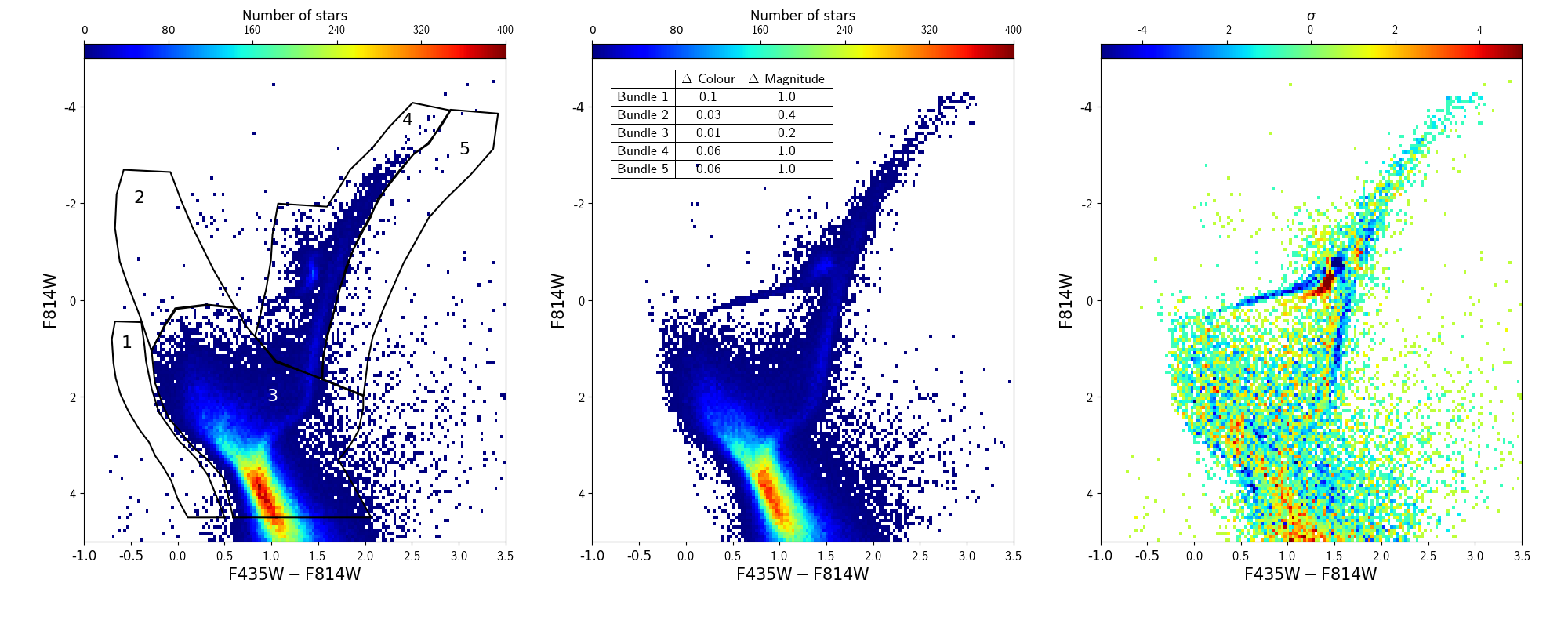} \\   
\caption{Colour-magnitude diagram fitting example and ``bundle'' strategy. From left to right: observed CMD (ACS field) with the bundles overplotted; ii) best fit CMD with a table showing the dimensions of the boxes in each bundle for the star counting; and iii) residual CMD (observed - best fit) in Poissonian sigmas.} 
\label{leoi_fitting_example} 
\end{figure*}

In this work, we recover SFHs by comparing observed and synthetic CMDs\footnote{Directly derived from stellar evolution models with the assumption of stellar population parameters and bolometric corrections} following the methodology carefully outlined in \citet[][]{2010ApJ...720.1225M}. In short, we define simple stellar populations (SSPs, using an age-metallicity grid) from synthetic CMDs to then obtain the combination of these SSPs that better reproduce observed diagrams. However, prior to the comparison itself there are a series of steps that need to be followed. First, we simulate observational effects (incompleteness and photometric errors) on the synthetic CMD based on the ASTs presented in Sect.~\ref{sec:reduction}. For this, we use a specifically-designed code named {\tt DisPar} (see Appendix~\ref{appendix2}). We then seek the linear combination  of SSPs that best fits the observed CMD. This is done using the code {\tt THESTORM} (``Tracing tHe Evolution of the STar fOrmation Rate and Metallicity), thoroughly described in \citet[][]{2015MNRAS.453L.113B} and \citet[][]{2018MNRAS.477.3507B}. {\tt THESTORM} compares the number of stars in colour-magnitude boxes of different sizes depending on their location within the observed and synthetic CMDs (a ``bundle'' strategy defines the colour-magnitude grid, with the size of the boxes in each bundle determining its importance in the fit, see Fig.~\ref{leoi_fitting_example}). In addition, {\tt THESTORM} carries out this comparison several times by applying shifts i) to the observed CMDs with respect to the synthetic CMD in colour and magnitude to assess possible uncertainties in the photometry, photometric zero points, distance, or reddening vector, anchoring the observed CMD to the position where the minimum $\chi^2$ is found ([$\Delta$(colour)$_{\rm min}$, $\Delta$(Mag)$_{\rm min}$]); ii) in the color-magnitude grid sampling the stars in the CMD; and iii) in the age-metallicity grid defining the SSP. The multiple SFHs calculated applying these shifts are combined in the final solution, and used in the calculation of the uncertainties in the Star Formation Rate (SFR) and Age-Metallicity Relation (AMR), following \citet[][]{2011ApJ...730...14H}.

Here we use synthetic CMDs\footnote{Synthetic CMDs are computed using an adapted version of the synthetic CMD code available at the BaSTI webpage (\url{http://basti.oa-teramo.inaf.it/index.html}). This code allows the user to obtain synthetic CMDs for a given photometric system and with flat distributions in age and metallicity.} computed using the BaSTI, solar scaled stellar evolution library \citep[][using bolometric corrections for the ACS and WFC3 passbands]{2004ApJ...612..168P}. They are composed by 50 million stars with ages between 0.03 and 14.0 Gyr and metallicities ranging from 0.0001 to 0.008 (covering the whole range of expected values for Leo~I). We also assumed a Kroupa initial mass fraction \citep[][]{2001MNRAS.322..231K}, an unresolved binary fraction of 50\% ($\beta$), a minimum mass ratio for binaries of 0.1 (q), and a Reimers mass loss parameter of 0.2 ($\eta$) for the computation of this synthetic CMD. Then, we divide the synthetic CMD according to the following basic age-metallicity grid to define the different SSPs:

Age: [0.03; 0.05; 0.1 - 1.0 in steps of 0.1; 1.2 - 2.0 in steps of 0.2; 2.5 - 6.0 in steps of 0.5; 7.0 - 14.0 in steps of 1] Gyr

Metallicity: [1, 2, 4, 7, 1.1, 1.6, 2.2, 2.9, 3.7, 4.6, 5.6, 6.7, 8.0] $\times$10$^{-4}$

The bundle strategy is depicted in Fig.~\ref{leoi_fitting_example} (left panel), superimposed in the observed CMD of Leo I, together best model (middle panel) and the residuals (right-hand panel). The bundles are designed to include in the fit most of the observed stars (i.e. most of the observed stars are within the ``bundles'') with at least 1 magnitude deeper than the location of the oMSTO (something that the photometric depth and accuracy of the analysed data allow). Small boxes (larger weight) are used in the bundles covering the better-modelled and denser MS region (bundle \#3), whereas larger boxes (less weight) are used in the bright MS (bundle \#2), RGB and RC regions (bundles \#4 and \#5, respectively) as well as a region bluer than the blue end of the MS (bundle \#1). The sizes of the boxes in each bundle are given in the table on the central panel of Fig.~\ref{leoi_fitting_example}, where the best fit CMD is represented. Finally, the right panel shows the residuals CMD. Note the quality of the fit, particularly in the MS region. Some differences in the area of the CMD dominated by helium-burning stars appear. These are due to two main effects: i) on the one hand, the BaSTI models adopted in this work use conductive opacity prescriptions that can be improved, slightly decreasing the brightness of HB stellar models \citep[][]{cassisi:07, Hidalgo2018}; ii) on the other hand, implementation of a variable mass loss efficiency during the RGB stage is preferable while computing synthetic CMDs \citep[][]{savino:19} rather than a fixed one as in this case ($\eta=0.2$, Reimers mass loss parameter). This has a small effect mainly in the colour distribution of stars along the core He-burning sequence. The low weight given during the fit to these conflict areas minimises its effect on the SFH recovery. During the course of this project we have also carefully tested the effect of the $\beta$ and q parameters defining the binary star population in the synthetic CMD, and that of the bundle strategy on the SFH recovery. Appendix~\ref{appendix1} summarises all these tests, legitimates the above described choice and expands on the adopted bundle strategy. The chosen approach is that named ``A2'' in appendix~\ref{appendix1}. The position of the [$\Delta$(colour)$_{\rm min}$, $\Delta$(Mag)$_{\rm min}$] was found at [0.03, 0.00] in the fitting of the ACS CMD.

\begin{figure}
\centering 
\includegraphics[width = 0.45\textwidth]{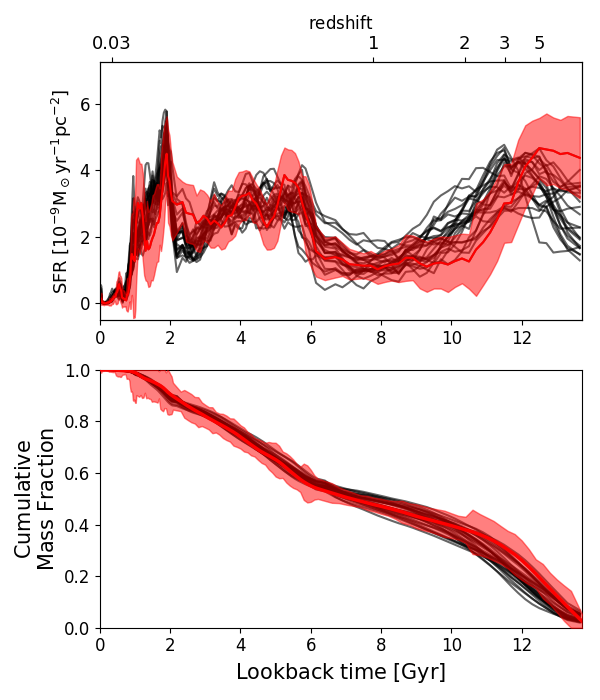} \\   
\caption{Global SFH recovered from the ACS field (central parts) of Leo~I. Top: star formation rate as a function of time. Bottom: cumulative mass fraction. In red we highlight the best solution (A2). Nevertheless, we show the solutions from all different tests (see Appendix~\ref{appendix1}) in black.} 
\label{leoi_sfh} 
\end{figure}

Fig.~\ref{leoi_sfh} shows the SFH derived for the region within Leo~I observed with the ACS, corresponding to the parameterization of the CMDs described above (red). For completeness, we also show all the solutions for all the tests (black, see Appendix~\ref{appendix1}). Note the robustness of the technique as all the solutions are compatible within errors. The main characteristics displayed by the Leo~I SFH are in common to all the tests, indicating that these results are independent on the details of the method. We find that Leo~I has experienced an extended SFH characterised by conspicuous star formation enhancements (0.4-0.65, 1.0-1.28, 1.65-2.5, 3.7-5.9, 10.5-13.7 Gyr ago). After the first prolonged star forming episode (older than $\sim$ 10 Gyr), an epoch of low star formation follows until $\sim$ 6 Gyr ago when star formation is heavily reignited. A more abrupt quenching of the star formation activity can be clearly observed around 1 Gyr ago, when the SFR is compatible with zero except for a well-defined burst 0.5 Gyr ago. The causes and interpretation of this peculiar behaviour will be discussed in Sect.~\ref{discussion}. 

The bottom panel of Fig.~\ref{leoi_sfh} shows the cumulative SFH, indicative of a nearly constant average SFH, except for the initial and final periods. Note that even though this kind of representation is useful for a broad characterisation of the evolution of the stellar mass content in a galaxy, it doesn't capture many rich and robust details of its SFH that can be associated to important events in its evolution, as will be discussed in Sect.~\ref{discussion}. 

\begin{figure}
\centering 
\includegraphics[width = 0.47\textwidth]{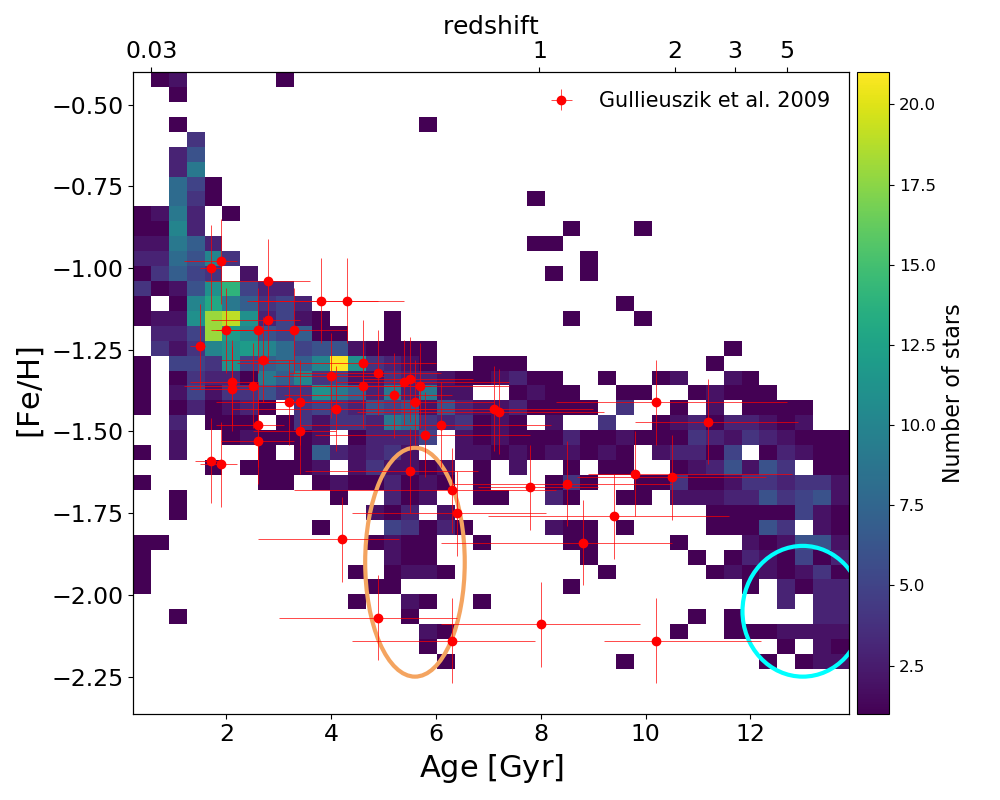} \\   
\caption{Global AMR recovered from the ACS field (central parts) of Leo~I (best solution, A2). On top of the solution presented in this work (coloured density map) we show the measurements from 54 RGB stars presented in Gullieuszik et al. 2009 (red points with errorbars). Ellipses highlight the tail of old, metal-poor stars (cyan) and that found around 5-6~Gyr ago (brown), interpreted as signs of a merger between Leo~I and an ultra-faint-like dwarf galaxy (see Sect.~\ref{discussion} for details).} 
\label{leoi_amr} 
\end{figure}

The recovered AMR is one of the main noveties of this work and is shown in Fig.~\ref{leoi_amr} as an age-metallicity density map. We observe chemical enrichment from [Fe/H] values close to $\sim$-1.9 for the oldest stars in Leo~I to $\sim$-0.75 for the youngest. We can distinguish two clear epochs of rapid chemical enrichment from $\sim$13 to 10 Gyr ago and from 2.5 to 1 Gyr. After that, the average stellar metallicity seems to slightly decrease (although this is an uncertain feature given the low number of stars younger than 1 Gyr found in the solution). For stars with intermediate-to-old ages ($\sim$2.5 to 10 Gyr) little or no chemical enrichment is found. Specially striking is the spread in metallicity shown by stars of ages older than 11 Gyr and between 5 and 6 Gyr, both coinciding with two important star forming periods. Another clear deviation from the general trend can be found at age $\sim$~9 Gyr and [Fe/H]~$\sim$~-0.9. However, whereas the first feature (metal-poor tail at ages $\sim$~5-6 and >~11~Gyr) can be observed in all tests shown in Fig.~\ref{leoi_sfh} and is considered reliable, this second one disappears in most of the cases and thus, it is not real but a natural fluctuation within the fitting procedure. Figure~\ref{leoi_amr} also compares the recovered AMR with the information on ages and metallicities of 54 RGB stars from the analysis of spectroscopic data in the \caii~triplet region \citep[][]{2009A&A...500..735G}. The resemblance between both approaches is remarkable, especially mimicking the low-metallicity tail at $\sim$~5-6 Gyr. The large age uncertainties affecting giant stars, together with the scarcity of stars (low SFR), makes the region between 7 to 10 Gyr more uncertain, but still compatible between both approaches.

\begin{figure}
\centering 
\includegraphics[width = 0.45\textwidth]{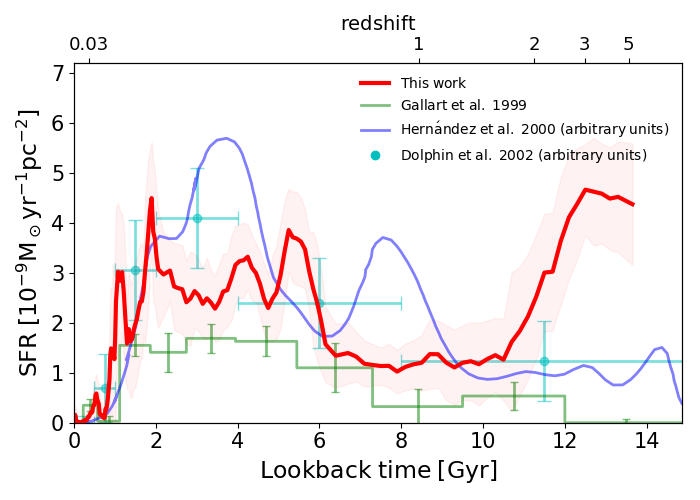} \\   
\caption{Comparison between SFH determinations from previous works. We compare the best solution (A2), with those presented in \citet[][green]{1999AJ....118.2245G}, \citet[][blue]{2000MNRAS.317..831H} and \citet[][]{Dolphin2002}, all using old data from WFPC2@HST. In the case of the \citet[][]{2000MNRAS.317..831H} and \citet[][]{Dolphin2002} determination the SFR has been normalised to arbitrary units.} 
\label{leoi_sfh_comparison} 
\end{figure}

This SFH is in broad agreement with the main features found in previous determinations \citep[e.g.][]{1999AJ....118.2245G, 2000MNRAS.317..831H, Dolphin2002}, and particularly with the presence of an important amount of intermediate-age population, the increase of the SFR occurred $\simeq$ 6 Gyr ago, and the quenching $\simeq$ 1 Gyr ago (see Fig.~\ref{leoi_sfh_comparison}). \citet{1999AJ....118.2245G} even recover the small star formation burst occurred $\simeq$ 0.5 Gyr ago. These previous studies, however, tend to find a substantially lower level of star formation at old ages, compared to what they find at intermediate ages, at odds with the present study. One might think that this could be partially due to the fact that the early determinations of the Leo I SFH were based on HST Wide Field Planetary Camera II (WFPC2) data which covered a central, smaller field of view than the current ACS data. However, as will be shown in Section~\ref{radial_leoi} below, we also find a high old SFR in the innermost parts of the analysed field. We conclude that it is the shallow CMD provided by the WFPC2 data that prevent previous works to fully unveil the old stellar populations (see Sect.~\ref{sec:data} for a comparison between both datasets). Interestingly, \citet[][]{2000MNRAS.317..831H} find a peak of star formation at $\sim$8~Gyr, where we find a sustained, low level of star formation (see Sect.~\ref{discussion} for a discussion on this). \citet[][]{Dolphin2002} claimed that the largest and more intense period of star formation expands from 3 to 1~Gyr ago, in agreement with two of our detected peaks. Overall, our SFH offers a much more detailed view of the evolution of Leo I with an unprecedented age resolution, including the metallicity evolution and the oldest population in Leo~I, which are poorly investigated in these previous studies due to the limited depth of the WFPC2 dataset.

\section{A radial view on the Leo~I evolution}
\label{radial_leoi}

Given the limited spatial coverage of our data, the SFH presented in Sect.~\ref{all_leoi} might not be representative of the Leo~I dwarf galaxy as a whole, especially in the presence of stellar population gradients in Leo~I. A wealth of observational evidence exists suggesting that the presence of such radial gradients are ubiquitous \citep[e.g.][]{2000ApJ...530L..85H, 2001AJ....122.3092H, 2004ApJ...617L.119T, 2006A&A...459..423B, 2012ApJ...761L..31B, 2012MNRAS.424.1113B, 2014A&A...570A..78B, 2015MNRAS.454.3996D, 2014MNRAS.444.3139M, 2016ApJ...829...86S, 2017MNRAS.467..208O}. In this sense, simulations are finally finding the presence of age gradients in simulated galaxies, with older stars dominating the outer parts \citet[e.g.][]{2019MNRAS.490.1186G}. As such, the SFH presented in Fig.~\ref{leoi_sfh} might be biased to young ages if similar gradients exist in Leo~I \citep[e.g.][]{2009A&A...500..735G, 2010MNRAS.404.1475H}, not providing a global view characteristic of this dwarf galaxy. The authors also suggest that the region around the half-light radius can be used as a good proxy for such global SFH. Fortunately, the analysed ACS data cover this region ($\sim$ 250 pc from its centre, \citealt[][]{2012AJ....144....4M}) and its quality allows us to study it separately (purpose of this section). In addition, a spatially resolved study of the Leo~I SFH can provide further insight on its build-up and evolution. 

\begin{figure*}
\centering 
\includegraphics[width = 0.95\textwidth]{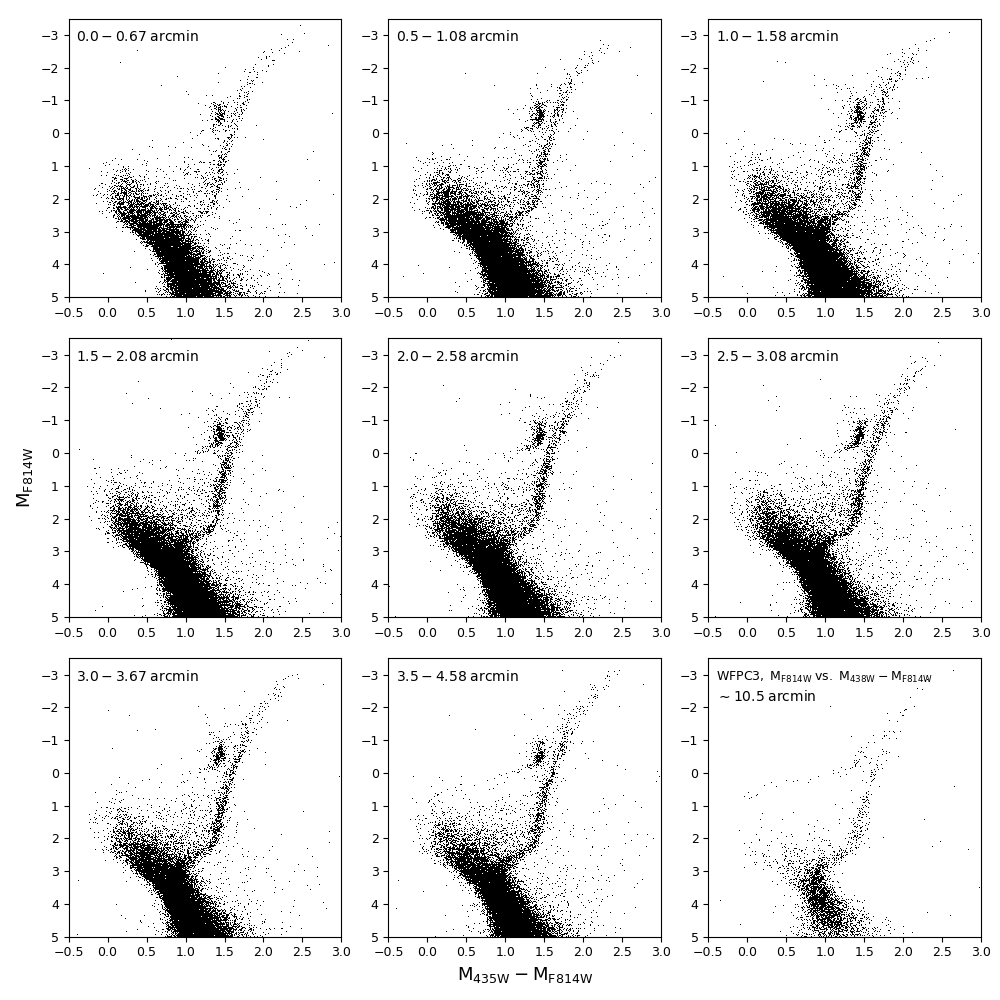} \\   
\caption{Observed colour-magnitude diagrams used in the analysis of the spatial variation of the Leo~I SFH. Galactocentric distance increases from the top-left to the bottom-right, being this last one the outermost, WFC3 parallel field. Colour-magnitude combinations are M$_{\rm F814W}$ and M$_{\rm F435W}$-M$_{\rm F814W}$ for the ACS fields and M$_{\rm F814W}$ and M$_{\rm F438W}$-M$_{\rm F814W}$ for the WFC3 field.} 
\label{leoi_radial_cmds} 
\end{figure*}

With this aim in mind, and benefiting from the exquisite quality of the analysed data (ACS and WFC3 parallel field, see Sect.~\ref{sec:data}), we split the ACS observed field into eight spatial regions. We adopted an elliptical isophotal binning scheme to define the different spatial regions under analysis. The elliptical apertures were centrered at 10$^{\rm h}$08$^{\rm m}$28$^{\rm s}$.1, +12$^{\rm o}$18$'$23$''$ \citep[J2000,][]{2012AJ....144....4M} with PA = 79$^{\rm o}$, $\epsilon$ = 0.21 \citep[][]{1995MNRAS.277.1354I} and partially overlap \citep[][]{2013ApJ...778..103H}. We preferred to use overlapping regions to increase the spatial sampling having a good number of stars in each aperture and to smooth out statistical fluctuations. The widths of the different apertures were varied accordingly as a compromise between the number of stars in each bin ($\sim$20000~-~30000) and the radial sampling of the ACS field. Finally, the most external region (region 9) is given by the location of the WFC3 parallel field and was analysed as a whole due to its low number of stars. These regions are overlaid to a DSS r-band image in Fig.~\ref{leoi_layout} and relevant information is given in Table~\ref{elliptical_tab}. All the observed CMDs for each spatial region can be found in Fig.~\ref{leoi_radial_cmds}.

\begin{table*}
{\normalsize
\centering
\begin{tabular}{cccccccc}
\hline\hline
Region & \multicolumn{2}{c}{Extension} & Number of stars & Area & Instrument & Average ($\sigma$) age & Average ($\sigma$) [Fe/H] \\ 
 &  (arcmin) & (pc) &  &  (pc$^2$) &   &(Gyr) & (dex) \\ 
$(1)$ & $(2)$ & $(3)$ & $(4)$ & $(5)$ & $(6)$ & $(7)$ & $(8)$\\ \hline
  1     &     0.0 - 0.67    &     0.0 -  49.4    &  17339  &  3702   &  ACS    & 6.8   (4.3) &  -1.54 (0.32) \\ 
  2     &     0.5 - 1.08    &    36.9 -  79.7    &  27668  &  6241   &  ACS    & 6.8   (4.4) &  -1.51 (0.34) \\ 
  3     &     1.0 - 1.58    &    73.8 - 116.6    &  32934  &  8144   &  ACS    & 6.9   (4.3) &  -1.51 (0.33) \\ 
  4     &     1.5 - 2.08    &   110.7 - 153.5    &  33411  &  9596   &  ACS    & 6.8   (4.3) &  -1.47 (0.34) \\ 
  5     &     2.0 - 2.58    &   147.6 - 190.4    &  33111  &  10628  &  ACS    & 6.8   (4.2) & -1.47  (0.34) \\ 
  6     &     2.5 - 3.08    &   184.5 - 227.3    &  31862  &  11546  &  ACS    & 7.0   (4.3) & -1.47  (0.36) \\ 
  7     &     3.0 - 3.67    &   221.4 - 270.8    &  25828  &  10107  &  ACS    & 7.3   (4.3) & -1.49  (0.38) \\ 
  8     &     3.5 - 4.58    &   258.3 - 338.0    &  24775  &  10945  &  ACS    & 7.6   (4.4) & -1.46  (0.40) \\ 
  9     &     $\sim$10.5    &    $\sim$771.8     &   3969  &  38920  &  WFC3   & 10.4  (3.4) & -1.67  (0.41) \\ \hline
\end{tabular}
\caption{Radial analysis of the Leo~I SFH. (1) Identifier of the radial region; (2) and (3) radial extension in arcmins and parsecs; (4) Number of stars within each region; (5) Area of the region (in pc$^2$); (6) Instrument; (7) Average (standard deviation) stellar age from the synthetic stars in the best fit model; (8) Average (standard deviation) stellar [Fe/H] from the best fit model.} 
\label{elliptical_tab}
}
\end{table*}

We apply the same methodology explained before to obtain the SFH from the nine observed radial CMDs. We follow the same bundle + synthetic CMD strategy (see Sect.~\ref{all_leoi} and Appendix~\ref{appendix1}) and we have anchored the solution to the [$\Delta$(colour)$_{\rm min}$, $\Delta$(Mag)$_{\rm min}$] found for the analysis of the whole Leo~I ACS CMD ([0.03, 0.00]). This restriction allows us to properly compare the average age and, especially, metallicity between different radial regions, such that the differences found are only attributable to real changes in their stellar content. Nine different ``dispersed'' synthetic CMDs (one for each region) are generated from the above described synthetic CMD after simulating errors. The ASTs used in this section have also been divided radially in the same eight apertures under analysis (ACS field) in order to be representative of the conditions in each radial bin. In the case of the parallel WFC3 field, it has been analysed as a whole (3969 stars after quality cuts in $\chi^2$ and {\it sharp} parameter are applied), using their own ASTs and a synthetic CMD computed for the WFC3 passbands.

\begin{figure}
\centering 
\includegraphics[width = 0.45\textwidth]{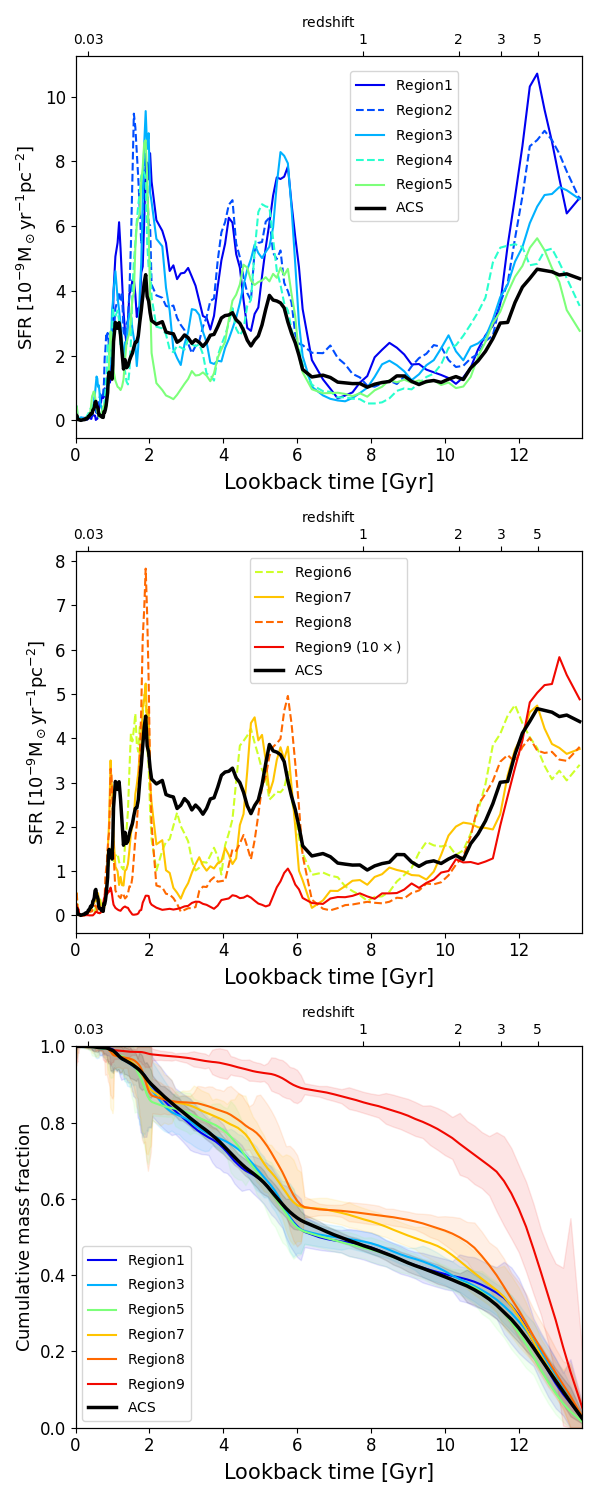} \\ 
\caption{Radial variation of the SFH in Leo~I. Left and middle panels show the star formation rate as a function of lookback time for the innermost and outermost regions, respectively. Right-hand panel displays the cumulative mass fractions for a set of radial apertures. As a reference, we have added to all plots the solution from the whole ACS field (A2 test, in black). Given the low star formation rate values found in region 9 we have applied a 10$\times$ factor for visualisation purposes.} 
\label{leoi_radial_sfh} 
\end{figure}

Figure~\ref{leoi_radial_sfh} shows the recovered SFH for the 9 spatial regions. A common pattern can be clearly observed in all regions in agreement to the global SFH features described in Sect.~\ref{all_leoi}. An initial epoch of intense star formation (continuing to 10 Gyr ago) is followed by a period of lower star forming activity, after which star formation is reignited, with prominent burst $\sim$5.5, 4, 2, 1 and 0.5 Gyr ago. However, some differences can be found when comparing the inner with the outer regions. Whereas in the inner $\sim$190~pc (regions 1-5, upper panel) such bursts coexist with an important background star formation, in the outer regions star formation is mainly concentrated in these star bursting events. This is also reflected in the cumulative representation of the SFH shown in the lower panel of Figure~\ref{leoi_radial_sfh}, where the lines corresponding to regions 6-9 depart from the almost straight line characteristic of the central Leo I SFH. It is particularly striking that the periods of enhanced star formation are found in all regions, {\it including the outermost region 9}, that is dominated by old stars ($\simeq$ 90\% and 80\% of star formation occurred prior to 6 Gyr ago and 10 Gyr ago, respectively).

\begin{figure*}
\centering 
\includegraphics[width = 0.95\textwidth]{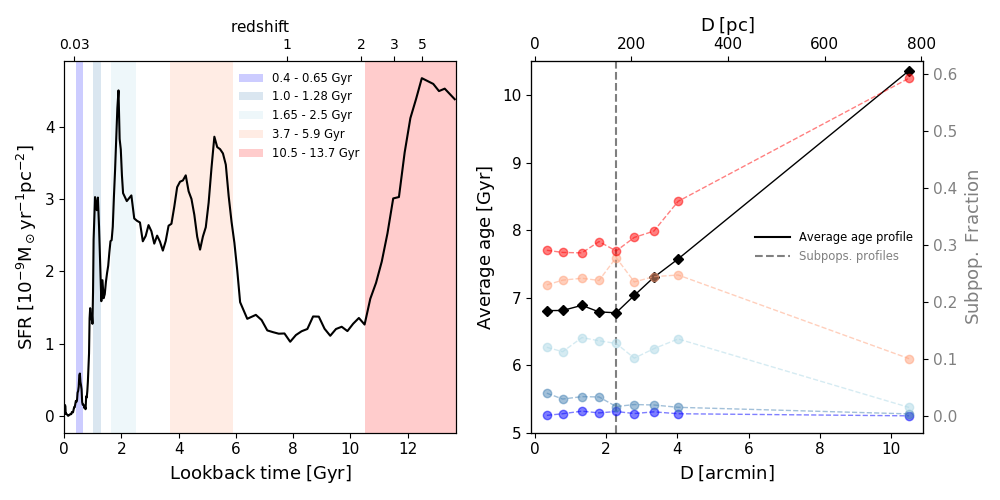} \\   
\caption{Stellar mean age radial profile and subpopulation spatial analysis. Left: Definition of the different stellar populations under analysis based on the SFH recovered using the whole ACS field (test A2). Right: We show the mean age profile (black, solid line; left hand y-axis) together with fraction of stars belonging to the different stellar subpopulations defined in the left-hand panel (coloured, dashed lines; right-hand y-axis).} 
\label{leoi_subpops} 
\end{figure*}

This dichotomy between inner and outer parts is clearly highlighted in Fig.~\ref{leoi_subpops} as well, which expands on the radial distribution of the stars associated with the above-defined star formation epochs. It also displays the Leo~I stellar mean age profile. Both the fraction of stars in each population within particular age ranges indicated in the left panel of Fig.~\ref{leoi_subpops} (coloured, dashed lines) and the age profile (black, solid line), display a flat trend up to $\sim$190 pc from the centre where they depart from it. Older average ages are found in these outermost regions mainly due to a decrease in the presence of stars younger than $\sim$4-6 Gyr and the lack of star formation in between bursts (i.e. 2 to 4~Gyr). These findings support the idea of an outside-in shrinking of the gaseous star forming disc \citep[][]{2009MNRAS.395.1455S} avoiding star formation in the outer parts in the last $\sim$4 to 6 Gyr, except maybe at the time of the previously described bursts (probably linked to violent events, see Sect.~\ref{discussion}). However, the possibility exists that \citep[as found in][]{2019MNRAS.490.1186G} outer stars with ages compatible with the reported bursts ($\sim$5.5, 4, 2, 1 and 0.5 Gyr) migrated to outer radii due to stellar feedback. Table~\ref{elliptical_tab} also shows the average stellar ages and [Fe/H] values (together with the standard deviations) of the stars populations regions 1 to 9 according to their corresponding best models. We see also here the above-mentioned dichotomy in the stellar age. The [Fe/H] values display an almost null gradient (considering the large dispersion of values), consistent with \citet[][]{2009A&A...500..735G}.

\section{Discussion}
\label{discussion}

\begin{figure*}
\centering 
\includegraphics[width = 0.63\textwidth]{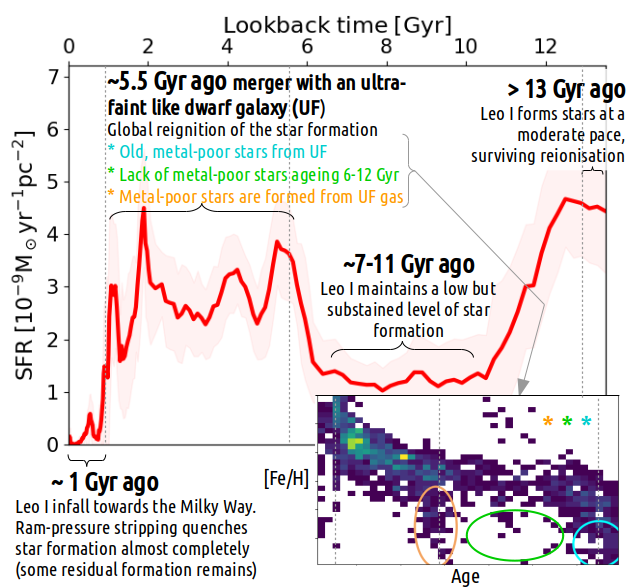} \\   
\caption{Scenario proposed for the temporal evolution of Leo~I based on the results presented in this work.} 
\label{leoi_scenario_info} 
\end{figure*}

The thorough characterisation of the stellar populations of Leo~I carried out in this work has further unveiled the complexity of this system, enabling an unprecedented understanding of its past. In this section, we will discuss the possible physical origin of the different features of the Leo~I SFH, from the oldest to the most recent ones, both in terms of the star formation and metallicity evolution. A graphical representation of the evolutionary scenario drawn from the results of this study is shown in Fig.~\ref{leoi_scenario_info}.

Reionisation has long been thought to play an important role in the early evolutionary stages of dwarf galaxies \citep[e.g.][]{1992MNRAS.256P..43E, 2000ApJ...542..535G}. In particular, it is expected to be able to regulate, and even truncate, star formation in low mass systems \citep[][]{2006MNRAS.371..401H, 2008MNRAS.390..920O}. Models predict that the UV background produced during reionisation \citep[occurring at redshift $\sim$7,][]{2016A&A...596A.108P} is capable of heating the gas and even driving it out of low-mass haloes, quenching star formation \citep[][]{2015MNRAS.450.4207B}. However, there is evidence indicating that reionisation alone cannot completely stop star formation in {\it classical} dwarf galaxies with stellar masses comparable with that of Leo~I  \citep[above $10^6$ M$_\odot$,][]{2010ApJ...720.1225M, 2010ApJ...722.1864M, 2011ApJ...730...14H}. On the contrary, the simultaneous combination of internal stellar feedback and reionization could completely halt star formation in dwarf galaxies \citep[][]{2015MNRAS.450.4207B}. We show here that Leo~I is a survivor of reionisation. The recovered SFH is compatible with a first peak of star formation occurring $\sim$12~Gyr ago, with lower (or constant) values of SFR previous to that main peak (see Fig.~\ref{leoi_sfh}), possibly indicating self-regulation at early times.  Whether the survival of Leo~I after reionisation was due to its already relatively high mass or to its formation in a relatively isolated environment with infrequent mergers is beyond the reach of this analysis.

\begin{figure}
\centering 
\includegraphics[width = 0.45\textwidth]{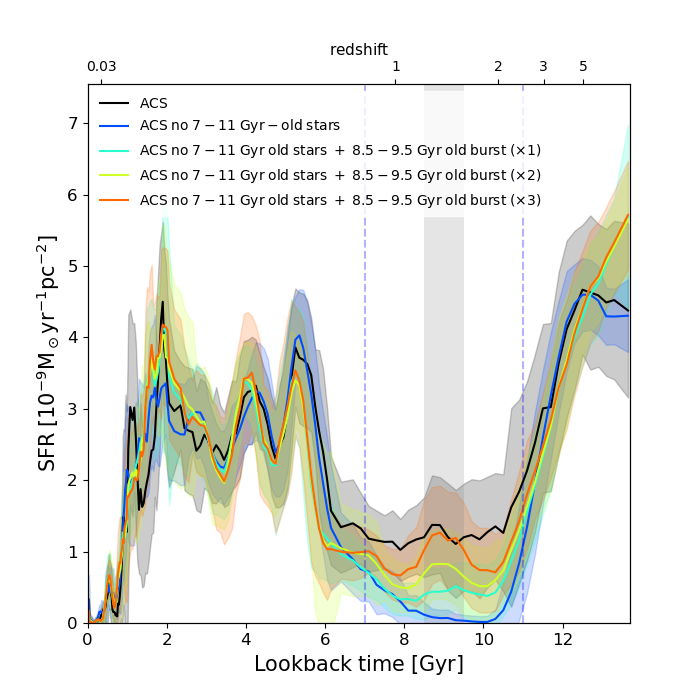} \\   
\caption{Characterisation of the residual star formation from $\sim$ 7 to 11 Gyr. We show the recovered SFH after analysing 4 mock CMDs: i) ACS solution removing all stars with ages in the range 7 to 11 Gyr old (blue); ii) Same as before but including stars in the range 8.5 to 9.5 Gyr (cyan), as well as twice and three times that number of stars (yellow and orange, respectively). For comparison we add the ACS solution (A2, black). We highlight here that none of the tests are able to reproduce the observed behaviour (constant star formation rate in this age range), suggesting that a low and near-constant star formation was maintained in Leo~I in this period.} 
\label{leoi_7-11Gyr_period} 
\end{figure}

After this initial phase in which Leo~I survived reionisation and the possible later peak in star formation ($\sim$12~Gyr ago), it entered a quiescent epoch (10 to 6~Gyr ago) characterised by the presence of a lower than average level of star formation. It is well-known that observational errors and the intrinsic limitations of the adopted methodology can affect our ability to recover well-defined bursts of star formation at old ages~\citep[][]{2020arXiv200209714R}. In addition, \citet[][]{2000MNRAS.317..831H} recover a conspicuous peak of star formation at $\sim$8~Gyr in this epoch where we find this quiescent period. \citet[][]{2020NatAs...4..965R} studied the ability of CMD-fitting techniques to recover star forming bursts in terms of time, duration, and age resolution. As a consequence, we have decided to further investigate the presence of the low level star formation in the 10 to 6~Gyr age range through a series of tests using mock stellar populations (see Fig.~\ref{leoi_7-11Gyr_period}). For that purpose, we have created synthetic CMDs mimicking both observational errors and the recovered SFH for the Leo~I ACS field (see Fig.~\ref{leoi_sfh}), adding some modifications in the $\sim$7 to 11~Gyr age range. In one test, we have removed all the stars in that range (see blue line in Fig.~\ref{leoi_7-11Gyr_period}). In the others, we added stars in the 8.5 to 9.5 Gyr range to mimic the presence of a burst of different strengths (green, lime, and orange lines). After applying our SFH recovery methodology to the CMDs of these mock stellar populations, we can conclude that we cannot reproduce the observed behaviour neither with a total absence of stars in the 7 to 11 Gyr range nor with the presence of a star forming burst within that range. These tests confirm that Leo~I kept a low, but sustained, level of star formation from $\sim$10~Gyr ago up to 6~Gyr ago. 

Star formation was reignited again at high level in Leo~I around 6~Gyr ago, nearly matching the level of SFR we measure 12~Gyr ago. There is an ample variety of processes that can result in the triggering of star formation in galactic (stellar instabilities, spiral waves, gas inflows and global perturbations) and local scales \citep[stellar pressure,][]{2012IAUS..284..317E}. In particular, galaxy mergers and interactions (both major and minor) are among the most efficient ones \citep[][]{kennicutt1987, mihos1994, hernquist1995, tissera2002, ellison2013}, and it has been shown that dwarf galaxies usually undergo one major and one minor merger between redshifts 5 and 0.5, triggering star formation and disrupting their morphology \citep[e.g.][]{2021MNRAS.500.4937M}. In this line, recent works are finding signatures of tidal interactions in the SFHs of Local Group dwarf galaxies \citep[][]{2020arXiv200209714R, 2020arXiv200307006M} and even in our own Galaxy \citep[][]{2020NatAs...4..965R}.

We suggest that, based on the SFH (both the SFR as a function of time and AMR) that we recover in our analysis (see Fig.~\ref{leoi_amr}), star formation was reignited as a consequence of a minor merger between Leo~I and another system $\sim$5.5~Gyr ago. Figure~\ref{leoi_amr} shows a large dispersion on the metallicity of stars born 5-6~Gyr ago (highlighted with a brown ellipse at intermediate ages), coinciding with the peak of star formation and characterised by a clear tail towards lower metallicities \citep[in agreement with the results from][]{2009A&A...500..735G}. Both, the high SFR as well as this metal-poor tail, can be explained if a merger between Leo~I and a lower mass stellar system (containing some gas) took place around that time. Such an event would have triggered star formation in Leo~I from its internally processed material (responsible for the bulk of stars in that age range), and could have re-ignited star formation in the merging system as well (producing the low-metallicity tail, owing to the mass-metallicity relation and the mass difference between both systems, e.g. \citealt[][]{2006ApJ...644..813E}; \citealt[][]{2016MNRAS.456.2140M}). This would naturally explain the wide distribution of stellar metallicities found in RR Lyrae stars too \citep[][]{2001ApJ...562L..39H}, also shown in the AMR presented in this work (see cyan ellipse in Fig.~\ref{leoi_amr}). In fact, Leo~I is the Milky Way satellite with (by far) the broadest distribution of stellar metallicity in RR Lyrae (old) stars (Mart\'inez-V\'azquez in prep.), and the reason for such broad distribution could be behind this merger. We hypothesise that the merging system could have been a ``gas-containing'' ultra-faint dwarf galaxy \citep[][]{2005AJ....129.2692W} similar to the rare Leo T \citep{Clementini2012LeoT, Weisz2012LeoT}. Such a system would have hosted some of the old, low-metallicity RR Lyrae stars that we see now in Leo~I, leading to the metal-poor tail reported in this work and the broad RR Lyrae stellar metallicity distribution \citep[][Mart\'inez-V\'azquez in prep.]{2001ApJ...562L..39H}. The scarcity of metal-poor stars with ages ranging from $\sim$11 to 6~Gyr old suggests the dominance of old stars in this ultra-faint-like system at the time it was accreted, with its star formation being reignited at the time of the interaction. In fact, a similar behaviour is observed in the SFH of Leo T \citep[][also Surot et al. in prep]{Clementini2012LeoT}, with a bursty SFR following a relatively low initial star formation activity. According to our SFH determination, at the time of the merge, Leo~I possessed half of the stellar mass it has now, resulting a mass-ratio between Leo~I and a typical ultra-faint dwarf of 15:1, enough to re-ignite important star formation in both systems \citep[][]{2020NatAs...4..965R}.

If this scenario was correct, one would expect that this event would have left some fingerprints not only in the chemical properties of the stars observed in Leo~I, but also kinematic (off-centred populations, comoving and chemical families of stars, etc.) and morphological features (although to a low extent given the mass ratio of the merger). Indeed, there is no clear evidence of the presence of such structures. However, there are intriguing findings in the literature that could be related. \citet[][]{2007ApJ...657..241K} found a subtle rise in the radial profile of the velocity dispersion of stars not associated with any kinematical substructure. \citet[][]{2008ApJ...675..201M} also claimed the existence of a significant velocity anisotropy for stars beyond 400'' ($\sim$ 500 pc) and linked it to a previous perigalacticon passage around the Milky Way ($\sim$1-2~Gyr ago). \citet[][]{2019ApJ...878..152L}, analysing data from \citet[][]{2007ApJ...657..241K} and \citet[][]{2007MNRAS.378..318B}, reported findings of a substructure composed by 6 stars ([Fe/H] from -1.78 to -1.52 dex) with low significance. In addition, we noticed some discrepancies in the exact location of the centre of this galaxy reported by different authors \citep[][]{2008ApJ...675..201M, 2012AJ....144....4M} as well as  slight (though not significant) differences in the spatial distribution of different stellar populations (selected in different positions of the ACS CMD). Given the low mass ratio and that signatures of such merging event might have been fully vanished by now, especially considering the close passage to the MW, we conclude that further and more detailed data are needed to properly assess this possibility. All evidence seems to support our ultra-faint-like merger scenario although we cannot discard a collision with a classical dwarf (see Fig.~\ref{leoi_scenario_info}. 

Finally, in agreement with previous determinations of the SFH of Leo~I \citep[][]{1999AJ....118.2245G, 2000MNRAS.317..831H}, star formation stopped $\sim$1~Gyr ago. This time is coincident with Leo~I's suggested pericentric passage to the Milky Way, as proposed from HST proper motion measurements \citep[][]{2013ApJ...768..139S} and more recently using Gaia data \citep[][]{2018A&A...616A..12G, 2018A&A...619A.103F}. We conclude that, after 5 Gyr of sustained, high level SFR, efficient ram pressure stripping from the Milky Way halo \citep{2015ApJ...808L..27W, 2016MNRAS.463.1916F} contributed to quench star formation $\sim$1~Gyr ago. Strikingly, our derived SFH suggests the presence of a small burst of star formation around 0.5~Gyr ago. Indeed, the careful analysis of the CMD of Leo~I (carried out in Sect.~\ref{leoi_cmd_desc}) shows the presence of stars as young as 100~Myr. These findings are in line to the recent work by \citet[][]{2019A&A...624A..11H}. The authors, using hydro-dynamical simulations, show that while ram-pressure efficiently strips hot and diffuse gas from the incoming dwarf systems, the cold gas may remain and be compressed, enhancing (a small tail) of star formation. Cold gas stripping will depend on the ratio between the thermal pressure and the ram pressure applied to the dwarf by the hot halo gas: if the thermal pressure is high, cold gas can be retained and even compressed to form new stars. In other words, the presence of stars younger than 1~Gyr (pericentric passage) in Leo~I suggests that the thermal pressure during the approach was high, implying a hot and dense Milky Way halo.

\vspace{3mm}

The radial analysis presented in Sect.~\ref{radial_leoi} adds another dimension to this study. On one hand, it shows that one must be cautious when extracting global conclusions based on a CMD not covering most of the analysed system, as spatial variations might bias the conclusions. \citet[][]{2018MNRAS.479.1514B} computed the total mass formed in stars by z~=~2 for a sample of 16 Local Group dwarf galaxies taking into account the fraction of old stars that was being missed in SFH studies due to their limited spatial coverage and presence of age gradients. In the same line, \citet[][]{2019MNRAS.490.1186G} suggested that the SFH recovered from a region around the half-light radius of a system would be representative of its global SFH. The similarities between what we call global SFH determined in Sect.~\ref{all_leoi} (see Fig.~\ref{leoi_sfh}) and that of region 7 in Sect.~\ref{radial_leoi} (see Fig.~\ref{leoi_radial_sfh}), located near the half light radius both support the findings by \citet[][]{2019MNRAS.490.1186G} and confirm the validity of the discussion so far. On the other hand, adding the spatial dimension to SFH studies poses important constraints to theoretical simulations on galaxy formation and evolution, allowing a better understanding on the physics governing dwarf galaxy evolution.

Figure~\ref{leoi_radial_sfh} and specially Fig.~\ref{leoi_subpops} show a clear dominance of progressively older populations as we move outwards in Leo~I \citep[in agreement with ][]{2000ApJ...530L..85H, 2009A&A...500..735G}. Strikingly, the inner $\sim$~190 pc of Leo~I seems to be fairly uniform, displaying very similar SFHs and the same census of stellar populations (see right hand panel of Fig.~\ref{leoi_subpops}). \citet[][]{2009MNRAS.395.1455S}, making use of smooth particle hydrodynamic simulations, found that elevated early star formation activity combined with supernova feedback could lead to: i) the shrinking of the gaseous star forming disc; and, ii) the ejection of stars to the outer parts. This would inevitably cause the appearance of positive age gradients and the accumulation of old stars in the outskirts of dwarf galaxies. The combination of both effects might well be the main reason behind the positive age gradient displayed by the outer regions of Leo~I, with different importance at different radius. Given the relative homogeneity of the SFHs in the inner parts of Leo~I, and the contrast between inner and outer parts in the last 5 Gyr, we could conclude that the gaseous star forming disc after the time of the last merger ($\sim$4.5 Gyr ago) had a radius of $\sim$190~pc. 

However, we find clear evidence suggesting that at least regions 5-8 were not totally depleted of gas after that early time, since conspicuous peaks of star formation are found in them at $\sim$13, 5.5, 2.0 and 1.0~Gyr ago. In these regions, the amount of star formation that occurred at these epochs is substantial, and clearly implies the presence of gas. However, its natural state and the physical conditions of the gas may have not allowed sustained star formation, and only when external effects came into play, star formation was triggered. That was the case 5.5~Gyr ago (proposed merger with an ultra faint-like dwarf system) and 1~Gyr ago (perigalacticon passage). The peak at 2~Gyr still remains a mystery. Nevertheless, the intensity of the 2-Gyr star forming peak, and its presence in the whole region under analysis (up to $\sim$750~pc) propose another possible event to be unveiled in the past history of Leo~I. The situation may be different in region 9, which may be representative of the far outskirts of Leo~I near its tidal radius. In this region, while enhancements in the SFR at $\sim$ 5.5, 2.0 and 1.0~Gyr ago are still noticeable, they are less prominent, and possibly linked to the ejection of stars to the outer parts via stellar feedback rather than merger-induced star formation \citep[][]{2019MNRAS.490.1186G}.

\section{Conclusions}
\label{conclusions}

In this work we have thoroughly analysed the deepest available HST CMDs of the Leo~I dwarf galaxy. These CMDs reach the oldest main sequence turnoffs with photometry of excellent quality, and cover a wide spatial extension of this galaxy. The combination of this dataset, together with state-of-the-art CMD fitting techniques and stellar evolutionary models, has allowed us to obtain a unprecedented detailed view of the past history of this interesting dwarf galaxy (see Fig.~\ref{leoi_scenario_info}). Our findings can be summarised as follows:

\begin{enumerate}
\item Leo~I has been able to maintain an extended and episodic star formation until $\sim$1~Gyr ago, when star formation quenched, possibly as a consequence of ram pressure stripping induced by the halo of the Milky Way after the latest pericentric passage. The absence of gas in Leo~I \citep[][]{1978AJ.....83..360K, 2009ApJ...696..385G}, together with its SFH may indicate that this dwarf galaxy has just finished its transition from dIrr to dSph. 
\item Conspicuous and recurrent peaks of star formation are found throughout Leo~I at $\sim$13, 5.5, 2.0 and 1.0~Gyr ago. The spatial extension of such bursts suggest a violent and external origin for recent ones, while the oldest one should correspond to the early phases of star formation in the system.
\item We find evidence of a wet merger $\sim$5.5~Gyr ago with an ultra-faint dwarf-like stellar system, responsible for such a star forming burst and the presence of extremely metal poor stars in Leo~I.
\item We hypothesise that another important event should have taken place $\sim$2~Gyr ago, giving rise to one of the most efficient star forming events in Leo~I.
\item Residual star formation in the last Gyr (after Leo~I's pericentric passage) suggest that cold gas survived ram pressure stripping, implying that the Milky Way halo was hot and dense at the moment of the encounter.
\item We find two clearly differentiated regions within Leo~I. The inner $\sim$190~pc present a quite homogeneous distribution of its stellar populations, displaying a very similar SFH. This may be the size of the star forming gaseous disc after the last large merger ($\sim$4.5 Gyr ago). The outer parts are characterised by a positive age gradient, with old populations dominating the outer parts, possibly as a consequence of the shrinking of the star forming gaseous disc and radial ejection of stars.

\end{enumerate}

Further studies as the one presented in this work, linking global with spatially-resolved SFHs, are of the utmost importance. They do not only permit an unprecedented characterisation of past events in the history of individual galaxies, they also allow to set further constraints on low-mass galaxy simulations and the modelling of baryonic physics. This will significantly improve our knowledge on how dwarf galaxies form and evolve.

\section*{Acknowledgements}

We thank the anonymous referee for useful comments improving the original version of this manuscript. TRL, MM, CG, TKF and GB acknowledge financial support through the grant AYA2017-89076-P (AEI/FEDER, UE) as well as by the Ministerio de Ciencia, Innovaci\'on y Universidades (MCIU), through the State Budget and by the Consejer\'\i a de Econom\'\i a, Industria, Comercio y Conocimiento of the Canary Islands Autonomous Community, through the Regional Budget (including IAC projects, ``The Local Group in multi-dimensions'' and TRACES). TRL acknowledges support from a Spinoza grant (NWO) awarded to A. Helmi. TRL is also supported by grant AYA2016-77237-C3-1-P (RAVET project), a MCIU Juan de la Cierva - Formaci\'on grant (FJCI-2016-30342) and acknowledges support from the Spanish Public State Employment Service (SEPE). This research has been partially supported by the Spanish projects PID2019-104928RB-I00 (MINECO/FEDER, UE) and ProID2017010132 (Gobierno de Canarias/FEDER, UE). SC acknowledges support from Premiale INAF ``MITIC'' and grant AYA2013-42781P from the Ministry of Economy and Competitiveness of Spain, he has also been supported by INFN (Iniziativa specifica TAsP). This research makes use of python (\url{http://www.python.org}); Matplotlib \citep[][]{hunter2007}, a suite of open-source python modules that provide a framework for creating scientific plots; and Astropy, a community-developed core Python package for Astronomy \citep[][]{astropy2013, 2018AJ....156..123A}.

\section*{Data Availability}

The main dataset analysed in this article are publicly available and can be retrieved from the HST archive webpage (\url{https://archive.stsci.edu/hst/}) under programs GO10520 and GO12270. Fully reduced photometry, artifitial star tests, synthetic CMDs, and star formation histories consequence of this analysis will be shared on reasonable request to the corresponding author.

%%%%%%%%%%%%%%%%%%%%%%%%%%%%%%%%%%%%%%%%%%%%%%%%%%



%\bibliographystyle{mnras}

\bibliographystyle{mnras}
\bibliography{bibliography}

\begin{thebibliography}{}
\makeatletter
\relax
\def\mn@urlcharsother{\let\do\@makeother \do\$\do\&\do\#\do\^\do\_\do\%\do\~}
\def\mn@doi{\begingroup\mn@urlcharsother \@ifnextchar [ {\mn@doi@}
  {\mn@doi@[]}}
\def\mn@doi@[#1]#2{\def\@tempa{#1}\ifx\@tempa\@empty \href
  {http://dx.doi.org/#2} {doi:#2}\else \href {http://dx.doi.org/#2} {#1}\fi
  \endgroup}
\def\mn@eprint#1#2{\mn@eprint@#1:#2::\@nil}
\def\mn@eprint@arXiv#1{\href {http://arxiv.org/abs/#1} {{\tt arXiv:#1}}}
\def\mn@eprint@dblp#1{\href {http://dblp.uni-trier.de/rec/bibtex/#1.xml}
  {dblp:#1}}
\def\mn@eprint@#1:#2:#3:#4\@nil{\def\@tempa {#1}\def\@tempb {#2}\def\@tempc
  {#3}\ifx \@tempc \@empty \let \@tempc \@tempb \let \@tempb \@tempa \fi \ifx
  \@tempb \@empty \def\@tempb {arXiv}\fi \@ifundefined
  {mn@eprint@\@tempb}{\@tempb:\@tempc}{\expandafter \expandafter \csname
  mn@eprint@\@tempb\endcsname \expandafter{\@tempc}}}

\bibitem[\protect\citeauthoryear{{Astropy Collaboration} et~al.,}{{Astropy
  Collaboration} et~al.}{2013}]{astropy2013}
{Astropy Collaboration} et~al., 2013, \mn@doi [\aap]
  {10.1051/0004-6361/201322068}, \href
  {http://adsabs.harvard.edu/abs/2013A&A...558A..33A} {558, A33}

\bibitem[\protect\citeauthoryear{{Astropy Collaboration} et~al.,}{{Astropy
  Collaboration} et~al.}{2018}]{2018AJ....156..123A}
{Astropy Collaboration} et~al., 2018, \mn@doi [\aj] {10.3847/1538-3881/aabc4f},
  \href {https://ui.adsabs.harvard.edu/abs/2018AJ....156..123A} {156, 123}

\bibitem[\protect\citeauthoryear{{Battaglia} et~al.,}{{Battaglia}
  et~al.}{2006}]{2006A&A...459..423B}
{Battaglia} G.,  et~al., 2006, \mn@doi [\aap] {10.1051/0004-6361:20065720},
  \href {https://ui.adsabs.harvard.edu/abs/2006A&A...459..423B} {459, 423}

\bibitem[\protect\citeauthoryear{{Battaglia}, {Rejkuba}, {Tolstoy}, {Irwin}  \&
  {Beccari}}{{Battaglia} et~al.}{2012a}]{2012MNRAS.424.1113B}
{Battaglia} G.,  {Rejkuba} M.,  {Tolstoy} E.,  {Irwin} M.~J.,   {Beccari} G.,
  2012a, \mn@doi [\mnras] {10.1111/j.1365-2966.2012.21286.x}, \href
  {https://ui.adsabs.harvard.edu/abs/2012MNRAS.424.1113B} {424, 1113}

\bibitem[\protect\citeauthoryear{{Battaglia}, {Irwin}, {Tolstoy}, {de Boer}  \&
  {Mateo}}{{Battaglia} et~al.}{2012b}]{2012ApJ...761L..31B}
{Battaglia} G.,  {Irwin} M.,  {Tolstoy} E.,  {de Boer} T.,   {Mateo} M.,
  2012b, \mn@doi [\apjl] {10.1088/2041-8205/761/2/L31}, \href
  {https://ui.adsabs.harvard.edu/abs/2012ApJ...761L..31B} {761, L31}

\bibitem[\protect\citeauthoryear{{Beccari} et~al.,}{{Beccari}
  et~al.}{2014}]{2014A&A...570A..78B}
{Beccari} G.,  et~al., 2014, \mn@doi [\aap] {10.1051/0004-6361/201424411},
  \href {https://ui.adsabs.harvard.edu/abs/2014A&A...570A..78B} {570, A78}

\bibitem[\protect\citeauthoryear{{Bellazzini}, {Gennari}, {Ferraro}  \&
  {Sollima}}{{Bellazzini} et~al.}{2004}]{2004MNRAS.354..708B}
{Bellazzini} M.,  {Gennari} N.,  {Ferraro} F.~R.,   {Sollima} A.,  2004,
  \mn@doi [\mnras] {10.1111/j.1365-2966.2004.08226.x}, \href
  {http://adsabs.harvard.edu/abs/2004MNRAS.354..708B} {354, 708}

\bibitem[\protect\citeauthoryear{{Belloni}, {Askar}, {Giersz}, {Kroupa}  \&
  {Rocha-Pinto}}{{Belloni} et~al.}{2017}]{2017MNRAS.471.2812B}
{Belloni} D.,  {Askar} A.,  {Giersz} M.,  {Kroupa} P.,   {Rocha-Pinto} H.~J.,
  2017, \mn@doi [\mnras] {10.1093/mnras/stx1763}, \href
  {http://adsabs.harvard.edu/abs/2017MNRAS.471.2812B} {471, 2812}

\bibitem[\protect\citeauthoryear{{Belokurov} et~al.,}{{Belokurov}
  et~al.}{2007}]{2007ApJ...654..897B}
{Belokurov} V.,  et~al., 2007, \mn@doi [\apj] {10.1086/509718}, \href
  {https://ui.adsabs.harvard.edu/abs/2007ApJ...654..897B} {654, 897}

\bibitem[\protect\citeauthoryear{{Ben{\'\i}tez-Llambay}, {Navarro}, {Abadi},
  {Gottl{\"o}ber}, {Yepes}, {Hoffman}  \& {Steinmetz}}{{Ben{\'\i}tez-Llambay}
  et~al.}{2015}]{2015MNRAS.450.4207B}
{Ben{\'\i}tez-Llambay} A.,  {Navarro} J.~F.,  {Abadi} M.~G.,  {Gottl{\"o}ber}
  S.,  {Yepes} G.,  {Hoffman} Y.,   {Steinmetz} M.,  2015, \mn@doi [\mnras]
  {10.1093/mnras/stv925}, \href
  {https://ui.adsabs.harvard.edu/abs/2015MNRAS.450.4207B} {450, 4207}

\bibitem[\protect\citeauthoryear{{Bermejo-Climent} et~al.,}{{Bermejo-Climent}
  et~al.}{2018}]{2018MNRAS.479.1514B}
{Bermejo-Climent} J.~R.,  et~al., 2018, \mn@doi [\mnras]
  {10.1093/mnras/sty1651}, \href
  {https://ui.adsabs.harvard.edu/abs/2018MNRAS.479.1514B} {479, 1514}

\bibitem[\protect\citeauthoryear{{Bernard} et~al.,}{{Bernard}
  et~al.}{2008}]{Bernard2008Tucana}
{Bernard} E.~J.,  et~al., 2008, \mn@doi [\apjl] {10.1086/588285}, \href
  {https://ui.adsabs.harvard.edu/abs/2008ApJ...678L..21B} {678, L21}

\bibitem[\protect\citeauthoryear{{Bernard}, {Ferguson}, {Chapman}, {Ibata},
  {Irwin}, {Lewis}  \& {McConnachie}}{{Bernard}
  et~al.}{2015}]{2015MNRAS.453L.113B}
{Bernard} E.~J.,  {Ferguson} A.~M.~N.,  {Chapman} S.~C.,  {Ibata} R.~A.,
  {Irwin} M.~J.,  {Lewis} G.~F.,   {McConnachie} A.~W.,  2015, \mn@doi [\mnras]
  {10.1093/mnrasl/slv116}, \href
  {http://adsabs.harvard.edu/abs/2015MNRAS.453L.113B} {453, L113}

\bibitem[\protect\citeauthoryear{{Bernard}, {Schultheis}, {Di Matteo}, {Hill},
  {Haywood}  \& {Calamida}}{{Bernard} et~al.}{2018}]{2018MNRAS.477.3507B}
{Bernard} E.~J.,  {Schultheis} M.,  {Di Matteo} P.,  {Hill} V.,  {Haywood} M.,
   {Calamida} A.,  2018, \mn@doi [\mnras] {10.1093/mnras/sty902}, \href
  {http://adsabs.harvard.edu/abs/2018MNRAS.477.3507B} {477, 3507}

\bibitem[\protect\citeauthoryear{{Bosler}, {Smecker-Hane}  \&
  {Stetson}}{{Bosler} et~al.}{2007}]{2007MNRAS.378..318B}
{Bosler} T.~L.,  {Smecker-Hane} T.~A.,   {Stetson} P.~B.,  2007, \mn@doi
  [\mnras] {10.1111/j.1365-2966.2007.11792.x}, \href
  {https://ui.adsabs.harvard.edu/abs/2007MNRAS.378..318B} {378, 318}

\bibitem[\protect\citeauthoryear{{Boylan-Kolchin}, {Bullock}  \&
  {Kaplinghat}}{{Boylan-Kolchin} et~al.}{2011}]{2011MNRAS.415L..40B}
{Boylan-Kolchin} M.,  {Bullock} J.~S.,   {Kaplinghat} M.,  2011, \mn@doi
  [\mnras] {10.1111/j.1745-3933.2011.01074.x}, \href
  {https://ui.adsabs.harvard.edu/abs/2011MNRAS.415L..40B} {415, L40}

\bibitem[\protect\citeauthoryear{{Cash}}{{Cash}}{1979}]{Cash1979}
{Cash} W.,  1979, \mn@doi [\apj] {10.1086/156922}, \href
  {https://ui.adsabs.harvard.edu/abs/1979ApJ...228..939C} {228, 939}

\bibitem[\protect\citeauthoryear{{Cassisi}, {Potekhin}, {Pietrinferni},
  {Catelan}  \& {Salaris}}{{Cassisi} et~al.}{2007}]{cassisi:07}
{Cassisi} S.,  {Potekhin} A.~Y.,  {Pietrinferni} A.,  {Catelan} M.,   {Salaris}
  M.,  2007, \mn@doi [\apj] {10.1086/516819}, \href
  {https://ui.adsabs.harvard.edu/abs/2007ApJ...661.1094C} {661, 1094}

\bibitem[\protect\citeauthoryear{{Clementini}, {Cignoni}, {Contreras Ramos},
  {Federici}, {Ripepi}, {Marconi}, {Tosi}  \& {Musella}}{{Clementini}
  et~al.}{2012}]{Clementini2012LeoT}
{Clementini} G.,  {Cignoni} M.,  {Contreras Ramos} R.,  {Federici} L.,
  {Ripepi} V.,  {Marconi} M.,  {Tosi} M.,   {Musella} I.,  2012, \mn@doi [\apj]
  {10.1088/0004-637X/756/2/108}, \href
  {https://ui.adsabs.harvard.edu/abs/2012ApJ...756..108C} {756, 108}

\bibitem[\protect\citeauthoryear{{Cole} et~al.,}{{Cole}
  et~al.}{2007}]{2007ApJ...659L..17C}
{Cole} A.~A.,  et~al., 2007, \mn@doi [\apjl] {10.1086/516711}, \href
  {https://ui.adsabs.harvard.edu/abs/2007ApJ...659L..17C} {659, L17}

\bibitem[\protect\citeauthoryear{{Cole}, {Weisz}, {Dolphin}, {Skillman},
  {McConnachie}, {Brooks}  \& {Leaman}}{{Cole}
  et~al.}{2014}]{2014ApJ...795...54C}
{Cole} A.~A.,  {Weisz} D.~R.,  {Dolphin} A.~E.,  {Skillman} E.~D.,
  {McConnachie} A.~W.,  {Brooks} A.~M.,   {Leaman} R.,  2014, \mn@doi [\apj]
  {10.1088/0004-637X/795/1/54}, \href
  {https://ui.adsabs.harvard.edu/abs/2014ApJ...795...54C} {795, 54}

\bibitem[\protect\citeauthoryear{{Dayal} \& {Ferrara}}{{Dayal} \&
  {Ferrara}}{2018}]{2018PhR...780....1D}
{Dayal} P.,  {Ferrara} A.,  2018, \mn@doi [\physrep]
  {10.1016/j.physrep.2018.10.002}, \href
  {https://ui.adsabs.harvard.edu/abs/2018PhR...780....1D} {780, 1}

\bibitem[\protect\citeauthoryear{{Di Cintio}, {Brook}, {Macci{\`o}}, {Stinson},
  {Knebe}, {Dutton}  \& {Wadsley}}{{Di Cintio}
  et~al.}{2014}]{2014MNRAS.437..415D}
{Di Cintio} A.,  {Brook} C.~B.,  {Macci{\`o}} A.~V.,  {Stinson} G.~S.,  {Knebe}
  A.,  {Dutton} A.~A.,   {Wadsley} J.,  2014, \mn@doi [\mnras]
  {10.1093/mnras/stt1891}, \href
  {https://ui.adsabs.harvard.edu/abs/2014MNRAS.437..415D} {437, 415}

\bibitem[\protect\citeauthoryear{{Dolphin}}{{Dolphin}}{2002}]{Dolphin2002}
{Dolphin} A.~E.,  2002, \mn@doi [\mnras] {10.1046/j.1365-8711.2002.05271.x},
  \href {https://ui.adsabs.harvard.edu/abs/2002MNRAS.332...91D} {332, 91}

\bibitem[\protect\citeauthoryear{{Dutton} et~al.,}{{Dutton}
  et~al.}{2016}]{2016MNRAS.461.2658D}
{Dutton} A.~A.,  et~al., 2016, \mn@doi [\mnras] {10.1093/mnras/stw1537}, \href
  {https://ui.adsabs.harvard.edu/abs/2016MNRAS.461.2658D} {461, 2658}

\bibitem[\protect\citeauthoryear{{Efstathiou}}{{Efstathiou}}{1992}]{1992MNRAS.256P..43E}
{Efstathiou} G.,  1992, \mn@doi [\mnras] {10.1093/mnras/256.1.43P}, \href
  {https://ui.adsabs.harvard.edu/abs/1992MNRAS.256P..43E} {256, 43P}

\bibitem[\protect\citeauthoryear{{Ellison}, {Mendel}, {Patton}  \&
  {Scudder}}{{Ellison} et~al.}{2013}]{ellison2013}
{Ellison} S.~L.,  {Mendel} J.~T.,  {Patton} D.~R.,   {Scudder} J.~M.,  2013,
  \mn@doi [\mnras] {10.1093/mnras/stt1562}, \href
  {https://ui.adsabs.harvard.edu/abs/2013MNRAS.435.3627E} {435, 3627}

\bibitem[\protect\citeauthoryear{{Elmegreen}}{{Elmegreen}}{2012}]{2012IAUS..284..317E}
{Elmegreen} B.~G.,  2012, in {Tuffs} R.~J.,  {Popescu} C.~C.,  eds,  IAU
  Symposium Vol. 284, The Spectral Energy Distribution of Galaxies - SED 2011.
  pp 317--329 (\mn@eprint {arXiv} {1201.3659}),
  \mn@doi{10.1017/S1743921312009350}

\bibitem[\protect\citeauthoryear{{Erb}, {Shapley}, {Pettini}, {Steidel},
  {Reddy}  \& {Adelberger}}{{Erb} et~al.}{2006}]{2006ApJ...644..813E}
{Erb} D.~K.,  {Shapley} A.~E.,  {Pettini} M.,  {Steidel} C.~C.,  {Reddy} N.~A.,
    {Adelberger} K.~L.,  2006, \mn@doi [\apj] {10.1086/503623}, \href
  {https://ui.adsabs.harvard.edu/abs/2006ApJ...644..813E} {644, 813}

\bibitem[\protect\citeauthoryear{{Faria}, {Feltzing}, {Lundstr{\"o}m},
  {Gilmore}, {Wahlgren}, {Ardeberg}  \& {Linde}}{{Faria}
  et~al.}{2007}]{2007A&A...465..357F}
{Faria} D.,  {Feltzing} S.,  {Lundstr{\"o}m} I.,  {Gilmore} G.,  {Wahlgren}
  G.~M.,  {Ardeberg} A.,   {Linde} P.,  2007, \mn@doi [\aap]
  {10.1051/0004-6361:20065244}, \href
  {https://ui.adsabs.harvard.edu/abs/2007A&A...465..357F} {465, 357}

\bibitem[\protect\citeauthoryear{{Fillingham}, {Cooper}, {Pace},
  {Boylan-Kolchin}, {Bullock}, {Garrison-Kimmel}  \& {Wheeler}}{{Fillingham}
  et~al.}{2016}]{2016MNRAS.463.1916F}
{Fillingham} S.~P.,  {Cooper} M.~C.,  {Pace} A.~B.,  {Boylan-Kolchin} M.,
  {Bullock} J.~S.,  {Garrison-Kimmel} S.,   {Wheeler} C.,  2016, \mn@doi
  [\mnras] {10.1093/mnras/stw2131}, \href
  {http://adsabs.harvard.edu/abs/2016MNRAS.463.1916F} {463, 1916}

\bibitem[\protect\citeauthoryear{{Fritz}, {Battaglia}, {Pawlowski},
  {Kallivayalil}, {van der Marel}, {Sohn}, {Brook}  \& {Besla}}{{Fritz}
  et~al.}{2018}]{2018A&A...619A.103F}
{Fritz} T.~K.,  {Battaglia} G.,  {Pawlowski} M.~S.,  {Kallivayalil} N.,  {van
  der Marel} R.,  {Sohn} S.~T.,  {Brook} C.,   {Besla} G.,  2018, \mn@doi
  [\aap] {10.1051/0004-6361/201833343}, \href
  {https://ui.adsabs.harvard.edu/abs/2018A&A...619A.103F} {619, A103}

\bibitem[\protect\citeauthoryear{{Gaia Collaboration} et~al.,}{{Gaia
  Collaboration} et~al.}{2018a}]{2018A&A...616A...1G}
{Gaia Collaboration} et~al., 2018a, \mn@doi [\aap]
  {10.1051/0004-6361/201833051}, \href
  {https://ui.adsabs.harvard.edu/abs/2018A&A...616A...1G} {616, A1}

\bibitem[\protect\citeauthoryear{{Gaia Collaboration} et~al.,}{{Gaia
  Collaboration} et~al.}{2018b}]{2018A&A...616A..12G}
{Gaia Collaboration} et~al., 2018b, \mn@doi [\aap]
  {10.1051/0004-6361/201832698}, \href
  {https://ui.adsabs.harvard.edu/abs/2018A&A...616A..12G} {616, A12}

\bibitem[\protect\citeauthoryear{{Gallart}, {Freedman}, {Aparicio}, {Bertelli}
  \& {Chiosi}}{{Gallart} et~al.}{1999a}]{1999AJ....118.2245G}
{Gallart} C.,  {Freedman} W.~L.,  {Aparicio} A.,  {Bertelli} G.,   {Chiosi} C.,
   1999a, \mn@doi [\aj] {10.1086/301078}, \href
  {http://adsabs.harvard.edu/abs/1999AJ....118.2245G} {118, 2245}

\bibitem[\protect\citeauthoryear{{Gallart} et~al.,}{{Gallart}
  et~al.}{1999b}]{1999ApJ...514..665G}
{Gallart} C.,  et~al., 1999b, \mn@doi [\apj] {10.1086/306967}, \href
  {https://ui.adsabs.harvard.edu/abs/1999ApJ...514..665G} {514, 665}

\bibitem[\protect\citeauthoryear{{Gallart} et~al.,}{{Gallart}
  et~al.}{2015}]{2015ApJ...811L..18G}
{Gallart} C.,  et~al., 2015, \mn@doi [\apjl] {10.1088/2041-8205/811/2/L18},
  \href {https://ui.adsabs.harvard.edu/abs/2015ApJ...811L..18G} {811, L18}

\bibitem[\protect\citeauthoryear{{Gnedin}}{{Gnedin}}{2000}]{2000ApJ...542..535G}
{Gnedin} N.~Y.,  2000, \mn@doi [\apj] {10.1086/317042}, \href
  {https://ui.adsabs.harvard.edu/abs/2000ApJ...542..535G} {542, 535}

\bibitem[\protect\citeauthoryear{{Graus} et~al.,}{{Graus}
  et~al.}{2019}]{2019MNRAS.490.1186G}
{Graus} A.~S.,  et~al., 2019, \mn@doi [\mnras] {10.1093/mnras/stz2649}, \href
  {https://ui.adsabs.harvard.edu/abs/2019MNRAS.490.1186G} {490, 1186}

\bibitem[\protect\citeauthoryear{{Grcevich} \& {Putman}}{{Grcevich} \&
  {Putman}}{2009}]{2009ApJ...696..385G}
{Grcevich} J.,  {Putman} M.~E.,  2009, \mn@doi [\apj]
  {10.1088/0004-637X/696/1/385}, \href
  {https://ui.adsabs.harvard.edu/abs/2009ApJ...696..385G} {696, 385}

\bibitem[\protect\citeauthoryear{{Grebel}}{{Grebel}}{1999}]{1999IAUS..192...17G}
{Grebel} E.~K.,  1999, in {Whitelock} P.,  {Cannon} R.,  eds,  IAU Symposium
  Vol. 192, The Stellar Content of Local Group Galaxies. p.~17 (\mn@eprint
  {arXiv} {astro-ph/9812443})

\bibitem[\protect\citeauthoryear{{Gullieuszik}, {Held}, {Saviane}  \&
  {Rizzi}}{{Gullieuszik} et~al.}{2009}]{2009A&A...500..735G}
{Gullieuszik} M.,  {Held} E.~V.,  {Saviane} I.,   {Rizzi} L.,  2009, \mn@doi
  [\aap] {10.1051/0004-6361/200811578}, \href
  {http://adsabs.harvard.edu/abs/2009A%26A...500..735G} {500, 735}

\bibitem[\protect\citeauthoryear{{Harbeck} et~al.,}{{Harbeck}
  et~al.}{2001}]{2001AJ....122.3092H}
{Harbeck} D.,  et~al., 2001, \mn@doi [\aj] {10.1086/324232}, \href
  {https://ui.adsabs.harvard.edu/abs/2001AJ....122.3092H} {122, 3092}

\bibitem[\protect\citeauthoryear{{Hausammann}, {Revaz}  \&
  {Jablonka}}{{Hausammann} et~al.}{2019}]{2019A&A...624A..11H}
{Hausammann} L.,  {Revaz} Y.,   {Jablonka} P.,  2019, \mn@doi [\aap]
  {10.1051/0004-6361/201834871}, \href
  {https://ui.adsabs.harvard.edu/abs/2019A&A...624A..11H} {624, A11}

\bibitem[\protect\citeauthoryear{{Held}, {Saviane}, {Momany}  \&
  {Carraro}}{{Held} et~al.}{2000}]{2000ApJ...530L..85H}
{Held} E.~V.,  {Saviane} I.,  {Momany} Y.,   {Carraro} G.,  2000, \mn@doi
  [\apjl] {10.1086/312505}, \href
  {https://ui.adsabs.harvard.edu/abs/2000ApJ...530L..85H} {530, L85}

\bibitem[\protect\citeauthoryear{{Held}, {Clementini}, {Rizzi}, {Momany},
  {Saviane}  \& {Di Fabrizio}}{{Held} et~al.}{2001}]{2001ApJ...562L..39H}
{Held} E.~V.,  {Clementini} G.,  {Rizzi} L.,  {Momany} Y.,  {Saviane} I.,   {Di
  Fabrizio} L.,  2001, \mn@doi [\apjl] {10.1086/338105}, \href
  {https://ui.adsabs.harvard.edu/abs/2001ApJ...562L..39H} {562, L39}

\bibitem[\protect\citeauthoryear{{Held}, {Gullieuszik}, {Rizzi}, {Girardi},
  {Marigo}  \& {Saviane}}{{Held} et~al.}{2010}]{2010MNRAS.404.1475H}
{Held} E.~V.,  {Gullieuszik} M.,  {Rizzi} L.,  {Girardi} L.,  {Marigo} P.,
  {Saviane} I.,  2010, \mn@doi [\mnras] {10.1111/j.1365-2966.2010.16358.x},
  \href {https://ui.adsabs.harvard.edu/abs/2010MNRAS.404.1475H} {404, 1475}

\bibitem[\protect\citeauthoryear{{Hernandez}, {Gilmore}  \&
  {Valls-Gabaud}}{{Hernandez} et~al.}{2000}]{2000MNRAS.317..831H}
{Hernandez} X.,  {Gilmore} G.,   {Valls-Gabaud} D.,  2000, \mn@doi [\mnras]
  {10.1046/j.1365-8711.2000.03809.x}, \href
  {https://ui.adsabs.harvard.edu/abs/2000MNRAS.317..831H} {317, 831}

\bibitem[\protect\citeauthoryear{{Hernquist} \& {Mihos}}{{Hernquist} \&
  {Mihos}}{1995}]{hernquist1995}
{Hernquist} L.,  {Mihos} J.~C.,  1995, \mn@doi [\apj] {10.1086/175940}, \href
  {https://ui.adsabs.harvard.edu/abs/1995ApJ...448...41H} {448, 41}

\bibitem[\protect\citeauthoryear{{Hidalgo} et~al.,}{{Hidalgo}
  et~al.}{2011}]{2011ApJ...730...14H}
{Hidalgo} S.~L.,  et~al., 2011, \mn@doi [\apj] {10.1088/0004-637X/730/1/14},
  \href {http://adsabs.harvard.edu/abs/2011ApJ...730...14H} {730, 14}

\bibitem[\protect\citeauthoryear{{Hidalgo} et~al.,}{{Hidalgo}
  et~al.}{2013}]{2013ApJ...778..103H}
{Hidalgo} S.~L.,  et~al., 2013, \mn@doi [\apj] {10.1088/0004-637X/778/2/103},
  \href {https://ui.adsabs.harvard.edu/abs/2013ApJ...778..103H} {778, 103}

\bibitem[\protect\citeauthoryear{{Hidalgo} et~al.,}{{Hidalgo}
  et~al.}{2018}]{Hidalgo2018}
{Hidalgo} S.~L.,  et~al., 2018, \mn@doi [\apj] {10.3847/1538-4357/aab158},
  \href {https://ui.adsabs.harvard.edu/abs/2018ApJ...856..125H} {856, 125}

\bibitem[\protect\citeauthoryear{{Hoeft}, {Yepes}, {Gottl{\"o}ber}  \&
  {Springel}}{{Hoeft} et~al.}{2006}]{2006MNRAS.371..401H}
{Hoeft} M.,  {Yepes} G.,  {Gottl{\"o}ber} S.,   {Springel} V.,  2006, \mn@doi
  [\mnras] {10.1111/j.1365-2966.2006.10678.x}, \href
  {https://ui.adsabs.harvard.edu/abs/2006MNRAS.371..401H} {371, 401}

\bibitem[\protect\citeauthoryear{Hunter}{Hunter}{2007}]{hunter2007}
Hunter J.~D.,  2007, Computing In Science \& Engineering, 9, 90

\bibitem[\protect\citeauthoryear{{Irwin} \& {Hatzidimitriou}}{{Irwin} \&
  {Hatzidimitriou}}{1995}]{1995MNRAS.277.1354I}
{Irwin} M.,  {Hatzidimitriou} D.,  1995, \mn@doi [\mnras]
  {10.1093/mnras/277.4.1354}, \href
  {http://adsabs.harvard.edu/abs/1995MNRAS.277.1354I} {277, 1354}

\bibitem[\protect\citeauthoryear{{Kennicutt}, {Keel}, {van der Hulst}, {Hummel}
   \& {Roettiger}}{{Kennicutt} et~al.}{1987}]{kennicutt1987}
{Kennicutt} Jr. R.~C.,  {Keel} W.~C.,  {van der Hulst} J.~M.,  {Hummel} E.,
  {Roettiger} K.~A.,  1987, \mn@doi [\aj] {10.1086/114384}, \href
  {https://ui.adsabs.harvard.edu/abs/1987AJ.....93.1011K} {93, 1011}

\bibitem[\protect\citeauthoryear{{Kirby} et~al.,}{{Kirby}
  et~al.}{2010}]{2010ApJS..191..352K}
{Kirby} E.~N.,  et~al., 2010, \mn@doi [\apjs] {10.1088/0067-0049/191/2/352},
  \href {http://adsabs.harvard.edu/abs/2010ApJS..191..352K} {191, 352}

\bibitem[\protect\citeauthoryear{{Kirby}, {Lanfranchi}, {Simon}, {Cohen}  \&
  {Guhathakurta}}{{Kirby} et~al.}{2011}]{2011ApJ...727...78K}
{Kirby} E.~N.,  {Lanfranchi} G.~A.,  {Simon} J.~D.,  {Cohen} J.~G.,
  {Guhathakurta} P.,  2011, \mn@doi [\apj] {10.1088/0004-637X/727/2/78}, \href
  {http://adsabs.harvard.edu/abs/2011ApJ...727...78K} {727, 78}

\bibitem[\protect\citeauthoryear{{Kirby}, {Cohen}, {Guhathakurta}, {Cheng},
  {Bullock}  \& {Gallazzi}}{{Kirby} et~al.}{2013}]{2013ApJ...779..102K}
{Kirby} E.~N.,  {Cohen} J.~G.,  {Guhathakurta} P.,  {Cheng} L.,  {Bullock}
  J.~S.,   {Gallazzi} A.,  2013, \mn@doi [\apj] {10.1088/0004-637X/779/2/102},
  \href {https://ui.adsabs.harvard.edu/abs/2013ApJ...779..102K} {779, 102}

\bibitem[\protect\citeauthoryear{{Knapp}, {Kerr}  \& {Bowers}}{{Knapp}
  et~al.}{1978}]{1978AJ.....83..360K}
{Knapp} G.~R.,  {Kerr} F.~J.,   {Bowers} P.~F.,  1978, \mn@doi [\aj]
  {10.1086/112211}, \href
  {https://ui.adsabs.harvard.edu/abs/1978AJ.....83..360K} {83, 360}

\bibitem[\protect\citeauthoryear{{Koch}, {Wilkinson}, {Kleyna}, {Gilmore},
  {Grebel}, {Mackey}, {Evans}  \& {Wyse}}{{Koch}
  et~al.}{2007}]{2007ApJ...657..241K}
{Koch} A.,  {Wilkinson} M.~I.,  {Kleyna} J.~T.,  {Gilmore} G.~F.,  {Grebel}
  E.~K.,  {Mackey} A.~D.,  {Evans} N.~W.,   {Wyse} R.~F.~G.,  2007, \mn@doi
  [\apj] {10.1086/510879}, \href
  {https://ui.adsabs.harvard.edu/abs/2007ApJ...657..241K} {657, 241}

\bibitem[\protect\citeauthoryear{{Kroupa}}{{Kroupa}}{2001}]{2001MNRAS.322..231K}
{Kroupa} P.,  2001, \mn@doi [\mnras] {10.1046/j.1365-8711.2001.04022.x}, \href
  {http://adsabs.harvard.edu/abs/2001MNRAS.322..231K} {322, 231}

\bibitem[\protect\citeauthoryear{{Lee}, {Freedman}, {Mateo}, {Thompson}, {Roth}
   \& {Ruiz}}{{Lee} et~al.}{1993}]{1993AJ....106.1420L}
{Lee} M.~G.,  {Freedman} W.,  {Mateo} M.,  {Thompson} I.,  {Roth} M.,   {Ruiz}
  M.-T.,  1993, \mn@doi [\aj] {10.1086/116736}, \href
  {https://ui.adsabs.harvard.edu/abs/1993AJ....106.1420L} {106, 1420}

\bibitem[\protect\citeauthoryear{{Lora}, {Grebel}, {Schmeja}  \& {Koch}}{{Lora}
  et~al.}{2019}]{2019ApJ...878..152L}
{Lora} V.,  {Grebel} E.~K.,  {Schmeja} S.,   {Koch} A.,  2019, \mn@doi [\apj]
  {10.3847/1538-4357/ab1b71}, \href
  {https://ui.adsabs.harvard.edu/abs/2019ApJ...878..152L} {878, 152}

\bibitem[\protect\citeauthoryear{{Ma}, {Hopkins}, {Faucher-Gigu{\`e}re},
  {Zolman}, {Muratov}, {Kere{\v{s}}}  \& {Quataert}}{{Ma}
  et~al.}{2016}]{2016MNRAS.456.2140M}
{Ma} X.,  {Hopkins} P.~F.,  {Faucher-Gigu{\`e}re} C.-A.,  {Zolman} N.,
  {Muratov} A.~L.,  {Kere{\v{s}}} D.,   {Quataert} E.,  2016, \mn@doi [\mnras]
  {10.1093/mnras/stv2659}, \href
  {https://ui.adsabs.harvard.edu/abs/2016MNRAS.456.2140M} {456, 2140}

\bibitem[\protect\citeauthoryear{{Martin} et~al.,}{{Martin}
  et~al.}{2021}]{2021MNRAS.500.4937M}
{Martin} G.,  et~al., 2021, \mn@doi [\mnras] {10.1093/mnras/staa3443}, \href
  {https://ui.adsabs.harvard.edu/abs/2021MNRAS.500.4937M} {500, 4937}

\bibitem[\protect\citeauthoryear{{Mart{\'\i}nez-V{\'a}zquez}
  et~al.,}{{Mart{\'\i}nez-V{\'a}zquez}
  et~al.}{2015}]{Martinez-Vazquez2015Sculptor}
{Mart{\'\i}nez-V{\'a}zquez} C.~E.,  et~al., 2015, \mn@doi [\mnras]
  {10.1093/mnras/stv2014}, \href
  {https://ui.adsabs.harvard.edu/abs/2015MNRAS.454.1509M} {454, 1509}

\bibitem[\protect\citeauthoryear{{Mateo}}{{Mateo}}{1998}]{1998ARA&A..36..435M}
{Mateo} M.~L.,  1998, \mn@doi [\araa] {10.1146/annurev.astro.36.1.435}, \href
  {https://ui.adsabs.harvard.edu/abs/1998ARA&A..36..435M} {36, 435}

\bibitem[\protect\citeauthoryear{{Mateo}, {Olszewski}  \& {Walker}}{{Mateo}
  et~al.}{2008}]{2008ApJ...675..201M}
{Mateo} M.,  {Olszewski} E.~W.,   {Walker} M.~G.,  2008, \mn@doi [\apj]
  {10.1086/522326}, \href
  {https://ui.adsabs.harvard.edu/abs/2008ApJ...675..201M} {675, 201}

\bibitem[\protect\citeauthoryear{{McConnachie}}{{McConnachie}}{2012}]{2012AJ....144....4M}
{McConnachie} A.~W.,  2012, \mn@doi [\aj] {10.1088/0004-6256/144/1/4}, \href
  {http://adsabs.harvard.edu/abs/2012AJ....144....4M} {144, 4}

\bibitem[\protect\citeauthoryear{{McMonigal} et~al.,}{{McMonigal}
  et~al.}{2014}]{2014MNRAS.444.3139M}
{McMonigal} B.,  et~al., 2014, \mn@doi [\mnras] {10.1093/mnras/stu1659}, \href
  {https://ui.adsabs.harvard.edu/abs/2014MNRAS.444.3139M} {444, 3139}

\bibitem[\protect\citeauthoryear{{Mihos} \& {Hernquist}}{{Mihos} \&
  {Hernquist}}{1994}]{mihos1994}
{Mihos} J.~C.,  {Hernquist} L.,  1994, \mn@doi [\apjl] {10.1086/187299}, \href
  {https://ui.adsabs.harvard.edu/abs/1994ApJ...425L..13M} {425, L13}

\bibitem[\protect\citeauthoryear{{Miyoshi} \& {Chiba}}{{Miyoshi} \&
  {Chiba}}{2020}]{2020arXiv200307006M}
{Miyoshi} T.,  {Chiba} M.,  2020, arXiv: 2003.07006, \href
  {https://ui.adsabs.harvard.edu/abs/2020arXiv200307006M} {p. arXiv:2003.07006}

\bibitem[\protect\citeauthoryear{{Monelli} et~al.,}{{Monelli}
  et~al.}{2010a}]{2010ApJ...720.1225M}
{Monelli} M.,  et~al., 2010a, \mn@doi [\apj] {10.1088/0004-637X/720/2/1225},
  \href {http://adsabs.harvard.edu/abs/2010ApJ...720.1225M} {720, 1225}

\bibitem[\protect\citeauthoryear{{Monelli} et~al.,}{{Monelli}
  et~al.}{2010b}]{2010ApJ...722.1864M}
{Monelli} M.,  et~al., 2010b, \mn@doi [\apj] {10.1088/0004-637X/722/2/1864},
  \href {https://ui.adsabs.harvard.edu/abs/2010ApJ...722.1864M} {722, 1864}

\bibitem[\protect\citeauthoryear{{Monelli} et~al.,}{{Monelli}
  et~al.}{2016}]{2016ApJ...819..147M}
{Monelli} M.,  et~al., 2016, \mn@doi [\apj] {10.3847/0004-637X/819/2/147},
  \href {https://ui.adsabs.harvard.edu/abs/2016ApJ...819..147M} {819, 147}

\bibitem[\protect\citeauthoryear{{Mu{\~n}oz}, {C{\^o}t{\'e}}, {Santana},
  {Geha}, {Simon}, {Oyarz{\'u}n}, {Stetson}  \& {Djorgovski}}{{Mu{\~n}oz}
  et~al.}{2018}]{2018ApJ...860...66M}
{Mu{\~n}oz} R.~R.,  {C{\^o}t{\'e}} P.,  {Santana} F.~A.,  {Geha} M.,  {Simon}
  J.~D.,  {Oyarz{\'u}n} G.~A.,  {Stetson} P.~B.,   {Djorgovski} S.~G.,  2018,
  \mn@doi [\apj] {10.3847/1538-4357/aac16b}, \href
  {https://ui.adsabs.harvard.edu/abs/2018ApJ...860...66M} {860, 66}

\bibitem[\protect\citeauthoryear{{Okamoto}, {Gao}  \& {Theuns}}{{Okamoto}
  et~al.}{2008}]{2008MNRAS.390..920O}
{Okamoto} T.,  {Gao} L.,   {Theuns} T.,  2008, \mn@doi [\mnras]
  {10.1111/j.1365-2966.2008.13830.x}, \href
  {https://ui.adsabs.harvard.edu/abs/2008MNRAS.390..920O} {390, 920}

\bibitem[\protect\citeauthoryear{{Okamoto}, {Arimoto}, {Tolstoy}, {Jablonka},
  {Irwin}, {Komiyama}, {Yamada}  \& {Onodera}}{{Okamoto}
  et~al.}{2017}]{2017MNRAS.467..208O}
{Okamoto} S.,  {Arimoto} N.,  {Tolstoy} E.,  {Jablonka} P.,  {Irwin} M.~J.,
  {Komiyama} Y.,  {Yamada} Y.,   {Onodera} M.,  2017, \mn@doi [\mnras]
  {10.1093/mnras/stx086}, \href
  {https://ui.adsabs.harvard.edu/abs/2017MNRAS.467..208O} {467, 208}

\bibitem[\protect\citeauthoryear{{Peebles}}{{Peebles}}{1965}]{1965ApJ...142.1317P}
{Peebles} P.~J.~E.,  1965, \mn@doi [\apj] {10.1086/148417}, \href
  {https://ui.adsabs.harvard.edu/abs/1965ApJ...142.1317P} {142, 1317}

\bibitem[\protect\citeauthoryear{{Pietrinferni}, {Cassisi}, {Salaris}  \&
  {Castelli}}{{Pietrinferni} et~al.}{2004}]{2004ApJ...612..168P}
{Pietrinferni} A.,  {Cassisi} S.,  {Salaris} M.,   {Castelli} F.,  2004,
  \mn@doi [\apj] {10.1086/422498}, \href
  {http://adsabs.harvard.edu/abs/2004ApJ...612..168P} {612, 168}

\bibitem[\protect\citeauthoryear{{Planck Collaboration} et~al.,}{{Planck
  Collaboration} et~al.}{2016}]{2016A&A...596A.108P}
{Planck Collaboration} et~al., 2016, \mn@doi [\aap]
  {10.1051/0004-6361/201628897}, \href
  {https://ui.adsabs.harvard.edu/abs/2016A&A...596A.108P} {596, A108}

\bibitem[\protect\citeauthoryear{{Ruiz-Lara} et~al.,}{{Ruiz-Lara}
  et~al.}{2018}]{2018A&A...617A..18R}
{Ruiz-Lara} T.,  et~al., 2018, \mn@doi [\aap] {10.1051/0004-6361/201732398},
  \href {http://adsabs.harvard.edu/abs/2018A%26A...617A..18R} {617, A18}

\bibitem[\protect\citeauthoryear{{Ruiz-Lara}, {Gallart}, {Bernard}  \&
  {Cassisi}}{{Ruiz-Lara} et~al.}{2020}]{2020NatAs...4..965R}
{Ruiz-Lara} T.,  {Gallart} C.,  {Bernard} E.~J.,   {Cassisi} S.,  2020, \mn@doi
  [Nature Astronomy] {10.1038/s41550-020-1097-0}, \href
  {https://ui.adsabs.harvard.edu/abs/2020NatAs...4..965R} {4, 965}

\bibitem[\protect\citeauthoryear{{Rusakov}, {Monelli}, {Gallart}, {Fritz},
  {Ruiz-Lara}, {Bernard}  \& {Cassisi}}{{Rusakov}
  et~al.}{2020}]{2020arXiv200209714R}
{Rusakov} V.,  {Monelli} M.,  {Gallart} C.,  {Fritz} T.~K.,  {Ruiz-Lara} T.,
  {Bernard} E.~J.,   {Cassisi} S.,  2020, arXiv: 2002.09714, \href
  {https://ui.adsabs.harvard.edu/abs/2020arXiv200209714R} {p. arXiv:2002.09714}

\bibitem[\protect\citeauthoryear{{Santana} et~al.,}{{Santana}
  et~al.}{2016}]{2016ApJ...829...86S}
{Santana} F.~A.,  et~al., 2016, \mn@doi [\apj] {10.3847/0004-637X/829/2/86},
  \href {https://ui.adsabs.harvard.edu/abs/2016ApJ...829...86S} {829, 86}

\bibitem[\protect\citeauthoryear{{Savino}, {Tolstoy}, {Salaris}, {Monelli}  \&
  {de Boer}}{{Savino} et~al.}{2019}]{savino:19}
{Savino} A.,  {Tolstoy} E.,  {Salaris} M.,  {Monelli} M.,   {de Boer} T.~J.~L.,
   2019, \mn@doi [\aap] {10.1051/0004-6361/201936077}, \href
  {https://ui.adsabs.harvard.edu/abs/2019A&A...630A.116S} {630, A116}

\bibitem[\protect\citeauthoryear{{Silk} \& {Mamon}}{{Silk} \&
  {Mamon}}{2012}]{2012RAA....12..917S}
{Silk} J.,  {Mamon} G.~A.,  2012, \mn@doi [Research in Astronomy and
  Astrophysics] {10.1088/1674-4527/12/8/004}, \href
  {https://ui.adsabs.harvard.edu/abs/2012RAA....12..917S} {12, 917}

\bibitem[\protect\citeauthoryear{{Sohn}, {Besla}, {van der Marel},
  {Boylan-Kolchin}, {Majewski}  \& {Bullock}}{{Sohn}
  et~al.}{2013}]{2013ApJ...768..139S}
{Sohn} S.~T.,  {Besla} G.,  {van der Marel} R.~P.,  {Boylan-Kolchin} M.,
  {Majewski} S.~R.,   {Bullock} J.~S.,  2013, \mn@doi [\apj]
  {10.1088/0004-637X/768/2/139}, \href
  {http://adsabs.harvard.edu/abs/2013ApJ...768..139S} {768, 139}

\bibitem[\protect\citeauthoryear{{Stetson}}{{Stetson}}{1987}]{stetson87}
{Stetson} P.~B.,  1987, \mn@doi [\pasp] {10.1086/131977}, \href
  {https://ui.adsabs.harvard.edu/abs/1987PASP...99..191S} {99, 191}

\bibitem[\protect\citeauthoryear{{Stetson}}{{Stetson}}{1994}]{stetson94}
{Stetson} P.~B.,  1994, \mn@doi [\pasp] {10.1086/133378}, \href
  {https://ui.adsabs.harvard.edu/abs/1994PASP..106..250S} {106, 250}

\bibitem[\protect\citeauthoryear{{Stinson}, {Dalcanton}, {Quinn}, {Gogarten},
  {Kaufmann}  \& {Wadsley}}{{Stinson} et~al.}{2009}]{2009MNRAS.395.1455S}
{Stinson} G.~S.,  {Dalcanton} J.~J.,  {Quinn} T.,  {Gogarten} S.~M.,
  {Kaufmann} T.,   {Wadsley} J.,  2009, \mn@doi [\mnras]
  {10.1111/j.1365-2966.2009.14555.x}, \href
  {https://ui.adsabs.harvard.edu/abs/2009MNRAS.395.1455S} {395, 1455}

\bibitem[\protect\citeauthoryear{{Tissera}, {Dom{\'{\i}}nguez-Tenreiro},
  {Scannapieco}  \& {S{\'a}iz}}{{Tissera} et~al.}{2002}]{tissera2002}
{Tissera} P.~B.,  {Dom{\'{\i}}nguez-Tenreiro} R.,  {Scannapieco} C.,
  {S{\'a}iz} A.,  2002, \mn@doi [\mnras] {10.1046/j.1365-8711.2002.05385.x},
  \href {https://ui.adsabs.harvard.edu/abs/2002MNRAS.333..327T} {333, 327}

\bibitem[\protect\citeauthoryear{{Tolstoy} et~al.,}{{Tolstoy}
  et~al.}{2004}]{2004ApJ...617L.119T}
{Tolstoy} E.,  et~al., 2004, \mn@doi [\apjl] {10.1086/427388}, \href
  {https://ui.adsabs.harvard.edu/abs/2004ApJ...617L.119T} {617, L119}

\bibitem[\protect\citeauthoryear{{Tolstoy}, {Hill}  \& {Tosi}}{{Tolstoy}
  et~al.}{2009}]{2009ARA&A..47..371T}
{Tolstoy} E.,  {Hill} V.,   {Tosi} M.,  2009, \mn@doi [\araa]
  {10.1146/annurev-astro-082708-101650}, \href
  {https://ui.adsabs.harvard.edu/abs/2009ARA&A..47..371T} {47, 371}

\bibitem[\protect\citeauthoryear{{Weisz} et~al.,}{{Weisz}
  et~al.}{2011}]{2011ApJ...739....5W}
{Weisz} D.~R.,  et~al., 2011, \mn@doi [\apj] {10.1088/0004-637X/739/1/5}, \href
  {https://ui.adsabs.harvard.edu/abs/2011ApJ...739....5W} {739, 5}

\bibitem[\protect\citeauthoryear{{Weisz} et~al.,}{{Weisz}
  et~al.}{2012}]{Weisz2012LeoT}
{Weisz} D.~R.,  et~al., 2012, \mn@doi [\apj] {10.1088/0004-637X/748/2/88},
  \href {https://ui.adsabs.harvard.edu/abs/2012ApJ...748...88W} {748, 88}

\bibitem[\protect\citeauthoryear{{Weisz}, {Dolphin}, {Skillman}, {Holtzman},
  {Gilbert}, {Dalcanton}  \& {Williams}}{{Weisz}
  et~al.}{2014a}]{2014ApJ...789..147W}
{Weisz} D.~R.,  {Dolphin} A.~E.,  {Skillman} E.~D.,  {Holtzman} J.,  {Gilbert}
  K.~M.,  {Dalcanton} J.~J.,   {Williams} B.~F.,  2014a, \mn@doi [\apj]
  {10.1088/0004-637X/789/2/147}, \href
  {https://ui.adsabs.harvard.edu/abs/2014ApJ...789..147W} {789, 147}

\bibitem[\protect\citeauthoryear{{Weisz}, {Dolphin}, {Skillman}, {Holtzman},
  {Gilbert}, {Dalcanton}  \& {Williams}}{{Weisz}
  et~al.}{2014b}]{2014ApJ...789..148W}
{Weisz} D.~R.,  {Dolphin} A.~E.,  {Skillman} E.~D.,  {Holtzman} J.,  {Gilbert}
  K.~M.,  {Dalcanton} J.~J.,   {Williams} B.~F.,  2014b, \mn@doi [\apj]
  {10.1088/0004-637X/789/2/148}, \href
  {https://ui.adsabs.harvard.edu/abs/2014ApJ...789..148W} {789, 148}

\bibitem[\protect\citeauthoryear{{Weisz}, {Dolphin}, {Skillman}, {Holtzman},
  {Gilbert}, {Dalcanton}  \& {Williams}}{{Weisz}
  et~al.}{2015}]{2015ApJ...804..136W}
{Weisz} D.~R.,  {Dolphin} A.~E.,  {Skillman} E.~D.,  {Holtzman} J.,  {Gilbert}
  K.~M.,  {Dalcanton} J.~J.,   {Williams} B.~F.,  2015, \mn@doi [\apj]
  {10.1088/0004-637X/804/2/136}, \href
  {https://ui.adsabs.harvard.edu/abs/2015ApJ...804..136W} {804, 136}

\bibitem[\protect\citeauthoryear{{Wetzel}, {Tollerud}  \& {Weisz}}{{Wetzel}
  et~al.}{2015}]{2015ApJ...808L..27W}
{Wetzel} A.~R.,  {Tollerud} E.~J.,   {Weisz} D.~R.,  2015, \mn@doi [\apjl]
  {10.1088/2041-8205/808/1/L27}, \href
  {http://adsabs.harvard.edu/abs/2015ApJ...808L..27W} {808, L27}

\bibitem[\protect\citeauthoryear{{Willman} et~al.,}{{Willman}
  et~al.}{2005}]{2005AJ....129.2692W}
{Willman} B.,  et~al., 2005, \mn@doi [\aj] {10.1086/430214}, \href
  {https://ui.adsabs.harvard.edu/abs/2005AJ....129.2692W} {129, 2692}

\bibitem[\protect\citeauthoryear{{del Pino}, {Aparicio}  \& {Hidalgo}}{{del
  Pino} et~al.}{2015}]{2015MNRAS.454.3996D}
{del Pino} A.,  {Aparicio} A.,   {Hidalgo} S.~L.,  2015, \mn@doi [\mnras]
  {10.1093/mnras/stv2174}, \href
  {https://ui.adsabs.harvard.edu/abs/2015MNRAS.454.3996D} {454, 3996}

\makeatother
\end{thebibliography}

\bsp

%%%%%%%%%%%%%%%%%%%%%%%%%%%%%%%%%%%%%%%%%%%%%%%%%%

%%%%%%%%%%%%%%%%% APPENDICES %%%%%%%%%%%%%%%%%%%%%

\appendix

\section{On the reliability of the recovered SFHs: Testing the robustness of the method:}
\label{appendix1}

The fine details of the recovered SFH of a stellar system via CMD fitting might be, in principle, influenced by the details of the used synthetic CMD as well as the adopted bundle strategy for the comparison. In this Appendix, we assess the effects of these choices on our results, and evaluate the robustness and reliability of our SFH determination by comparing a number of realisations exploring different input parameters. Due to the large number of stars within the ACS field, we decided to carry out these tests using the CMD from this field (containing 191,756 stars). 

\begin{figure*}
\centering 
\includegraphics[width = 0.75\textwidth]{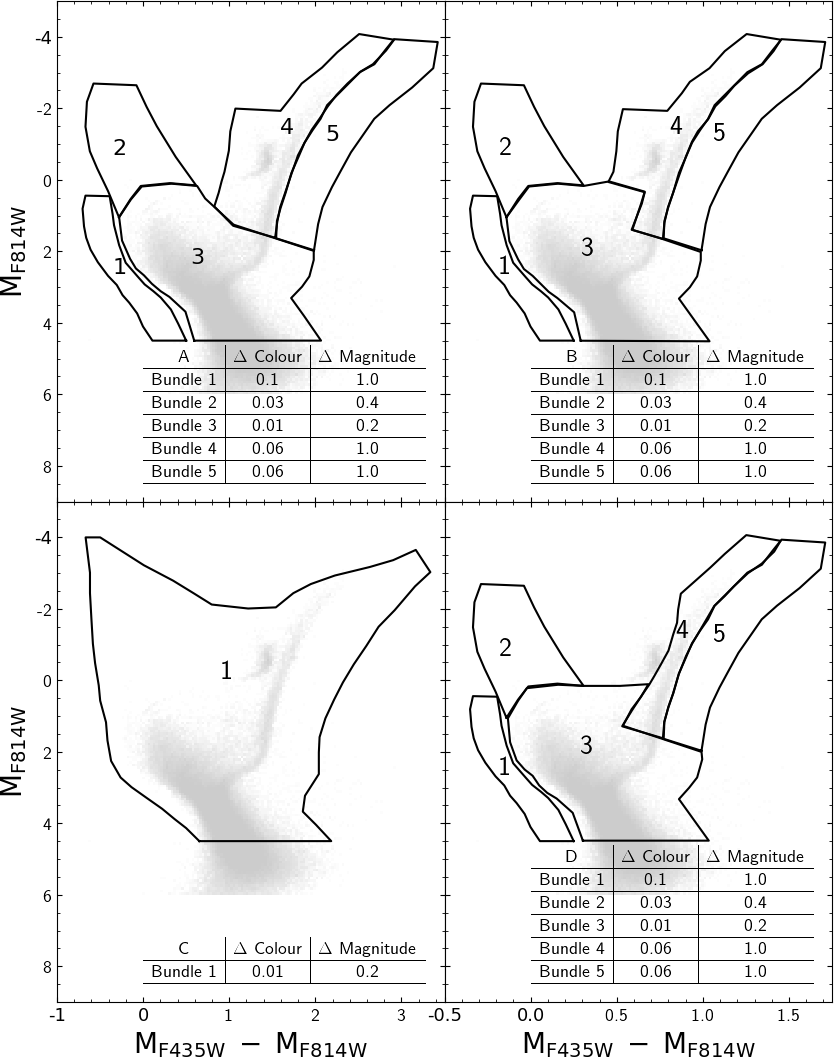} \\   
\caption{SFH recovery tests, bundle strategy. This figure depicts the different bundle strategies examined during this work. The inset tables indicate the dimensions of the boxes in which the different bundles are divided for the star counting (see Sect.~\ref{all_leoi}).} 
\label{bundles} 
\end{figure*}

All the tests and their main peculiarities are described in Table~\ref{recovery_tests_tab} and Fig.~\ref{bundles}. In short, six different synthetic diagrams \citep[all using the BaSTI isochrones,][]{2004ApJ...612..168P} are computed assuming different unresolved binary fractions ($\beta$) and minimum mass ratios for binaries (q, denoted by number from 1 to 6). In addition, four different bundle strategies have been used for each of the six synthetic CMDs, covering most of the stars with M$_{F814W}$ < 4.5 (denoted by letters from A to D). In total, 24 different tests are compared to single out the optimal strategy. 

\begin{figure}
\centering 
\includegraphics[width = 0.43\textwidth]{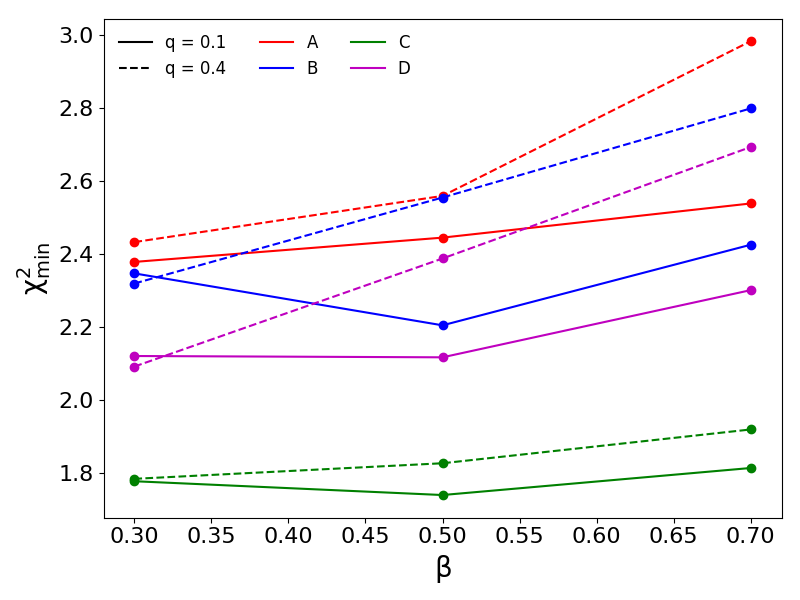} \\   
\caption{Variation of $\chi^2$ as a function of the unresolved binary fraction ($\beta$) for different minimum binary mass ratios (solid lines for q~=~0.1 and dashed lines for q~=~0.4) and different bundle strategies (red for A, blue for B, green for C, and magenta for D).} 
\label{q_beta_plot} 
\end{figure}

Binary stars are common in the Galactic field as well as in local dwarf galaxies. Thus, the way we incorporate them in the synthetic CMDs might affect the final SFH recovery \citep[][]{1999AJ....118.2245G, 2017MNRAS.471.2812B}. For this reason, we test different unresolved binary fractions ($\beta$ of 0.3, 0.5, and 0.7) and different minimum mass ratios (q of 0.1, allowing for stars of very different masses in the pairs; and q of 0.4, restricting to stars of similar mass in the pair). Figure~\ref{q_beta_plot} shows the $\chi^2$ of the 24 different tests distinguishing between different binary populations (line styles and x-axis) and bundle strategy (colours). In general, lower values of $\chi^2$ are found in the case of q~=~0.1 and $\beta$ of 0.3-0.5 (with 0.5 being slightly preferred). Careful examination of the fit residuals in the colour-magnitude plane show the presence of a clear, doubled MS when q~=~0.4 is used, suggesting that Leo~I hosts binaries with members of very different masses and arguing against this particular value. These tests clearly show that configuration number 2, q~=~0.1 and $\beta$~=~0.5, is preferred.

\begin{table*}
\begin{center}
\begin{tabular}{ccc|cl}
\hline\hline
 Synthetic &  &  & Bundle &  \\  
CMD test & q & $\beta$ & strategy & Bundle description \\ \hline
  {\bf 1}     &     0.1    &     0.3  & {\bf A} & MS in small boxes, sub-Giant+RGB+RC in big boxes \\ 
  {\bf 2}     &     0.1    &     0.5  & {\bf B} & similar to A + sub-Giant in small boxes \\ 
  {\bf 3}     &     0.1    &     0.7  & {\bf C} & whole CMD in small boxes \\ 
  {\bf 4}     &     0.4    &     0.3  & {\bf D} & similar to B, but avoiding the RC \\
  {\bf 5}     &     0.4    &     0.5  &   &  \\ 
  {\bf 6}     &     0.4    &     0.7  &   &  \\  \hline
\end{tabular}
\caption{SFH recovery tests. The 24 different realizations analysed during these tests are the combinations of i) 6 different unresolved binary fraction ($\beta$) and minimum mass ratio for binaries (q) as well as ii) 4 different bundle strategies regarding the inclusion and sampling of the Red clump (RC), Red Giant Branch (RGB), and sub-Giant region (see also Fig.~\ref{bundles}).} 
\end{center}
\label{recovery_tests_tab}
\end{table*}

In the adopted methodology, the bundle strategy determines the regions of the diagram from which information is extracted to recover the SFH. In general, one might think that avoiding regions of the CMD containing stars in less-understood and advanced stellar evolutionary phases or regions suffering from higher age-metallicity degeneracy should be favoured. Tests A to D expands on this. Strategy A represents an approach in which most of the CMD is considered for the fit, with the MS having more importance than other regions. In the cases B and D we include the sub-giant within the MS bundle and fit or avoid the red clump (RC) area. Strategy C adopts a single bundle well-sampled with the use of small boxes. 

\begin{figure}
\centering 
\includegraphics[width = 0.43\textwidth]{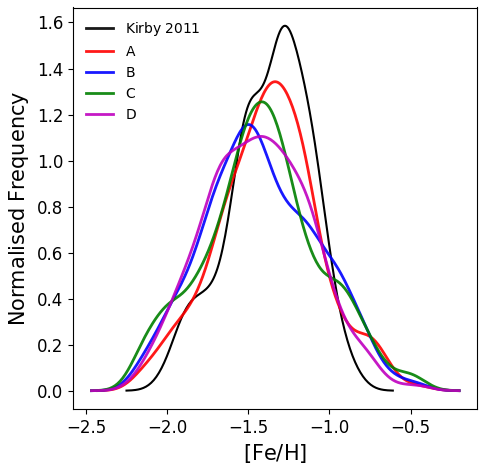} \\   
\caption{[Fe/H] distribution of RGB stars within the ACS field of view from \citet[][black line]{2011ApJ...727...78K} and from different bundle strategies using synthetic CMD number 2 (red for A, blue for B, green for C, and magenta for D).} 
\label{RGB_distr} 
\end{figure}

A direct comparison of the different $\chi^2$ values recovered to determine the best bundle strategy (as done before) would not be appropriate. {\tt THESTORM} ultimately relies on the minimisation of the $\chi^2$ function described in \citet[][]{Cash1979}:
\begin{equation}
\chi^{2}=2\sum_{j} (M^{j}-O^{j}\ln{M}^{j})
\end{equation}
where {\it M} and {\it O} would correspond to the model and observed CMDs, respectively, and {\it j} would run over all the boxes defined. As such, all the colour-magnitude boxes in which we divide the CMDs for the fit will contribute to the final $\chi^2$, affecting its value (see tables in Fig.~\ref{bundles}, and Sect.~\ref{all_leoi}). As a consequence, given the different number of boxes among strategies and CMD coverage (see Fig.~\ref{bundles}) we cannot simply compare $\chi^2$ values to assess the suitability of one strategy over the other. For this reason, and following a criterion previously used in \citet[][]{2018A&A...617A..18R}, the choice of the bundle strategy is made based on the comparison between the metallicity distributions (RGB region) that we find in each attempt with that observed for RGB stars by \citealt[][]{2011ApJ...727...78K} \citep[see also][]{2010ApJS..191..352K}, based on spectral synthesis of iron absorption lines. Fig.~\ref{RGB_distr} compares the [Fe/H] distribution of RGB stars within the ACS field of view determined by \citet[][]{2011ApJ...727...78K} and the distributions inferred using the different strategies (using always synthetic CMD \#2). We should clarify at this point that, for the comparison, we have used only RGB stars in the same range of magnitudes as that covered by \citet[][]{2011ApJ...727...78K}. Although all distributions are fairly compatible (especially considering the difficulty of reproducing the exact metallicity distribution of RGB stars from CMD fitting), configuration A is the one that results in a RGB metallicity distribution that better resembles the one found by \citet[][]{2011ApJ...727...78K}.

Adding the bundle strategy information to the one regarding the binary population we conclude that, among all the tests, A2 is preferred over the rest. Thus, all the analysis presented in the main body of the paper is based on this choice. As can be seen in Fig.~\ref{leoi_sfh} (black lines), all different tests are compatible within errors and the same different star formation bursts defined in Sect.~\ref{all_leoi} are found in all tests, highlighting the robustness and stability of this implementation of the CMD fitting technique. The analysis presented in this Appendix allows us to claim that the results presented in this work are not contingent on the particular set of input parameters used in the recovery of SFHs.

% DisPar

\section{{\tt DisPar}: Simulating observational effects on synthetic Colour-Magnitude Diagrams}
\label{appendix2}

\begin{figure}
\centering 
\includegraphics[width = 0.43\textwidth]{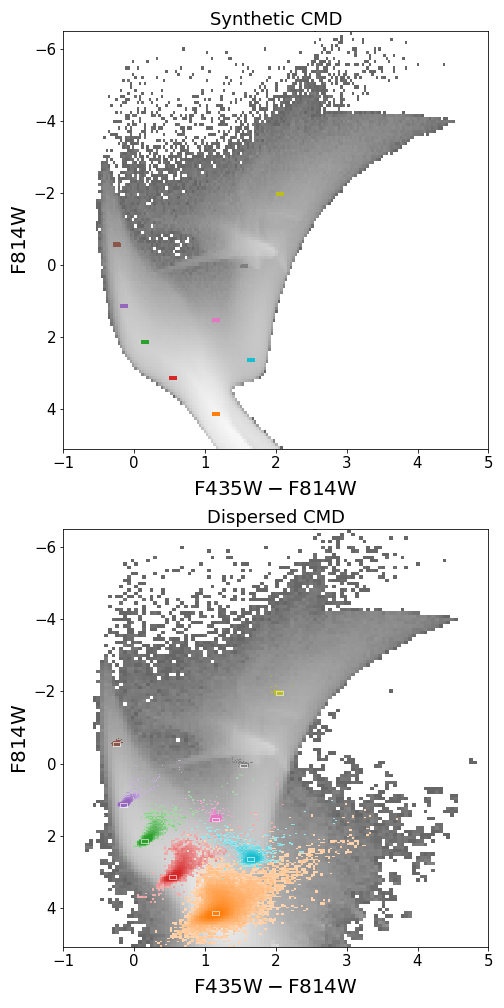} \\   
\caption{Error simulation with {\tt DisPar}. Left-hand panel: synthetic CMD number 2 before error simulation, some particular regions are highlighted in different colours. Right-hand panel: same synthetic CMD after simulating photometric errors taking into account artificial star tests (see Sect.~\ref{sec:reduction}) using {\tt DisPar}. The coloured density maps represent the position of the synthetic stars highlighted in the left-hand panel after the error simulation. As expected, fainter stars are more affected by crowding and photometric errors than brighter stars.} 
\label{DisPar_fig} 
\end{figure}

Before comparing the observed CMDs (from which we want to derive the SFH) with the set of SSPs extracted from the synthetic CMD (as described in Sect.~\ref{all_leoi}), it is necessary to use the information on incompleteness and photometric errors affecting the observed CMDs (and derived from the ASTs, see Sect.~\ref{sec:reduction}, as a function of both position in the CMD and physical position in the sky) to simulate uncertainties in the error-free synthetic CMDs. This error simulation is a key step towards the determination of SFHs as observational effects are able to affect the actual colours and magnitudes of observed stars, measuring colour and magnitude values far from the real ones (especially important at faint magnitudes below the oldest MSTO, see Fig.~\ref{DisPar_fig} and \citealt[][]{1999AJ....118.2245G}). The simulation of observational effects is done using a specifically designed code named {\tt DisPar}. In what follows we provide a short description on the practical procedure as well as the technical details behind {\tt DisPar}.

{\tt DisPar} goes through each and every star in the synthetic CMD of magnitudes $(M1_s, M2_s)$ and position $(X_s,Y_s)$\footnote{For explanatory purposes we will generalise to magnitudes $M1$ and $M2$ (generic filters $1$ and $2$) and positions $X$ and $Y$. The spatial distribution of observed stars in the sky (X,Y) is simulated in the synthetic CMD as well. In the particular case of this paper we use filters $F435W$ and $F814W$ (ACS); and $F438W$ and $F814W$ (WFC3). For the positions in the sky we use pixels (image coordinates). Sky coordinates (RA,DEC) could be used as well.}. For each synthetic star it considers all artificial stars from the ASTs within user-defined colour-magnitude ($[col,mag]=[0.06,0.1]$ in this particular case) and position boxes ($[X, Y]=[200px,200px]$) with input and recovered magnitudes $(M1_i, M2_i)$ and $(M1_r, M2_r)$, respectively, and input position $(X_i,Y_i)$. If a minimum number of artificial stars is found within the boxes (user-defined, 10 in this case), one is randomly selected and the magnitudes of the synthetic stars under consideration are modified accordingly, $M1_s'=M1_s-(M1_i-M1_r)$ and $M2_s'=M2_s-(M2_i-M2_r)$. In the case that this minimum number of artificial stars requirement is not fulfilled, the size of the boxes is doubled and the search starts again. If the synthetic star has no counterpart in the ASTs, its magnitudes are not modified and the star is flagged in the output file containing the dispersed CMD. An example of {\tt DisPar} applied to synthetic CMD number 2 is shown in Fig.~\ref{DisPar_fig}.

{\tt DisPar} algorithm basically performs searches in an efficient way. To this end, it relies on two key factors:
\begin{itemize}
\item the use of a data structure that allows to perform quick searches,
\item and the parallelization of the code.
\end{itemize}
Note that, if the number of synthetic stars is $m$ and the number of artificial stars is $n$, the complexity of a straight forward search is $O(mn)$\footnote{The so called big $O$ notation is used in computer science to classify algorithms according to how their run time requirements (or \textit{time complexity}) grow as the input size grows. An algorithm is said to be $O(f(n))$ if its time complexity doesn't grow faster than function $f(n)$, being $n$ the input size.} since, in the worst case, for any of the $m$ synthetic stars, all the $n$ artificial stars must be checked to look for the defined number of similar ones. This complexity can be reduced by using some data structure that eases the searches, such as AVL trees. An AVL tree is a binary search tree in which the difference between the heights of the left and right subtrees of any node is less than or equal to one. On such structure, the insertion, deletion, rebalancing, and searching operations can all be performed in $O(\log n)$, being $n$ the data size. 

When comparing the synthetic and artificial stars, four criteria are taken into account (position in the CMD, position in the sky). To this end, we created a custom data structure consisting of four nested AVL trees, one for each searching criteria. Moreover, for the sake of computational efficiency, the resulting dynamic data structure is then 'flattened', that is, each nested AVL tree is stored in an array, preserving relations between trees. As a result of these improvements, the computational complexity of {\tt DisPar} is set to $O(m \log n)$.

Finally, {\tt DisPar} was coded in C/C++ in three different ways, using parallel programming. The source code of the different implementations, as well as the description of the input files and parameters, and a user manual (so far only in Spanish, currently working on the English version), are available at the public repository \url{https://bitbucket.org/ULLfypIAC/workspace/projects/FYP}.

The {\tt DisPar} version used in this paper runs on multicore CPUs using OpenMP. Another version allows to choose between running on Nvidia GPUs using the CUDA programming model, or on multicore CPUs using OpenMP. Both versions can be found in folder {\tt crowding-effects-cli-iterative} and receive as input plain text files. The third version of the code uses also OpenMP and allows to modify the set of searching criteria, at the cost of loosing some computational speed. This code receives as input a JSON file, and it is placed in folder {\tt crowding-effects-cli-recursive}.

\begin{figure}
\centering 
\includegraphics[width = 0.45\textwidth]{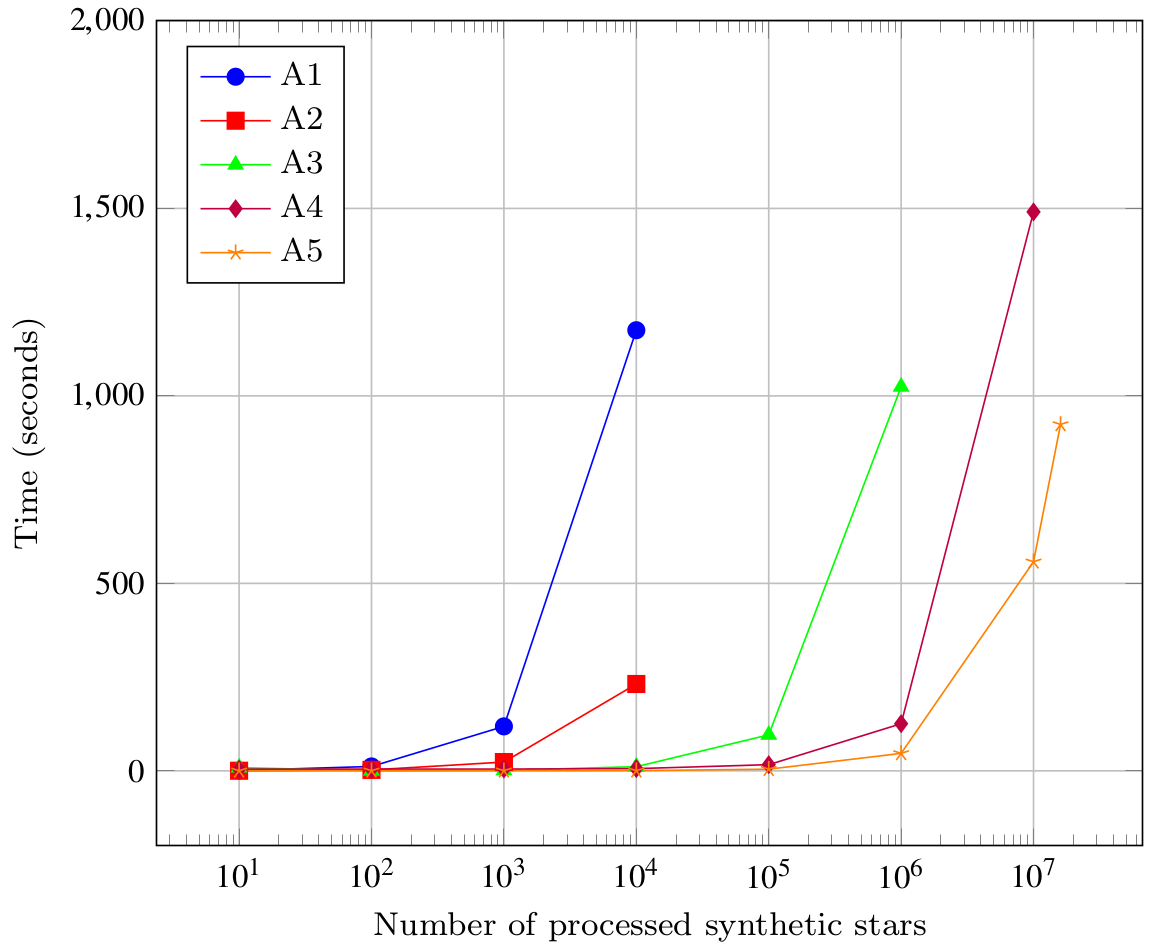} \\   
\caption{Evolution of the computing times for the different dispersion algorithms.} 
\label{DisPar_fig_comp_times} 
\end{figure}

To evaluate the performance of {\tt DisPar} we conducted several computational experiments on a data set with nearly 12000000 artificial stars and a number of synthetic stars ranging from 10 to 16000000. All the experiments were carried out on a PC under Kubuntu with a processor Intel Core i7-4930K at 3.4 GHz, 32 GB of RAM, and graphic card Nvidia GeForce Titan Black, GPU GK110, at 967 MHZ, with 2880 CUDA cores and a 6 GB VRAM. A time limit of 2000 seconds was imposed to each run. The results are displayed in Table~\ref{tab:leoi_dispar_comp_times}. Column \textit{nStars} shows the number of synthetics stars used in each experiment. The remaining columns show computing times in seconds for the following dispersion algorithms:
\begin{itemize}
  \item \textit{A1}: Sequential $O(mn)$ basic algorithm.
  \item \textit{A2}: $O(mn)$ basic algorithm parallelized using OpenMP.
  \item \textit{A3}: {\tt DisPar} OpenMP algorithm with possibility of choosing the set of searching criteria.
  \item \textit{A4}: {\tt DisPar} OpenMP algorithm with fixed number of searching criteria (4).
  \item \textit{A5}: {\tt DisPar} CUDA algorithm with fixed number of searching criteria (4).
\end{itemize}
A blank space in the table means that the algorithm has failed to fulfil the dispersion problem within the time limit. Fig. \ref{DisPar_fig_comp_times} shows the evolution of the computing times of the different algorithms as the number of synthetic stars increases. These results show that {\tt DisPar} clearly outperforms the basic dispersion algorithm, and that it is able cope with the dispersion problem for a large number of stars at a reasonable computational effort.

\begin{table*}
{\normalsize
\centering
\begin{tabular}{rrrrrr}
\hline\hline
   nStars       &       A1          &    A2     &         A3    &          A4  &           A5 \\ 
   \hline
       10   &       1.1139       &   0.2796     &     8.5115    &      4.5686   &      0.1332 \\
      100    &     11.6606       &   2.5325     &     0.1615    &      4.7227   &      0.2156 \\
     1000    &    118.5543       &  23.4513     &     1.3283    &      4.7142   &      0.33   \\
    10000    &   1175.1146       & 231.5997     &    11.1537    &      5.8857   &     0.464  \\
   100000    &          --       &       --     &    96.0288    &     16.4766   &      4.4536 \\
  1000000    &          --       &       --     &    1024.29    &    125.6757   &      46.5718 \\
 10000000    &          --       &       --     &         --    &   1490.7665   &   557.6948  \\
 16000000   &           --       &       --     &         --    &          --   &    923.8403 \\
\hline
\end{tabular}
\caption{Computing times for the different dispersion algorithms.} 
\label{tab:leoi_dispar_comp_times}
}
\end{table*}

%%%%%%%%%%%%%%%%%%%%%%%%%%%%%%%%%%%%%%%%%%%%%%%%%%

% Don't change these lines
\bsp	% typesetting comment
\label{lastpage}
\end{document}